\documentclass[lettersize,journal]{IEEEtran}
\usepackage{amsmath, amsfonts}
\usepackage{array}
\usepackage[caption=false, font=normalsize, labelfont=sf, textfont=sf]{subfig}
\usepackage{textcomp}
\usepackage{stfloats}
\usepackage{url}
\usepackage{verbatim}
\usepackage{graphicx}
\usepackage{cite}
\usepackage{orcidlink}

\usepackage{enumitem}
\usepackage{multirow}
\usepackage{graphicx}
\usepackage[labelfont=bf]{caption}
\usepackage{subcaption}
\usepackage{color}
\usepackage{tabu}
\usepackage{booktabs}
\usepackage{tabularx, booktabs}

\makeatletter
\newif\if@restonecol
\makeatother

\usepackage[linesnumbered, ruled, vlined]{algorithm2e}
\usepackage{algpseudocode}

\newcolumntype{C}[1]{>{\centering\let\newline\\\arraybackslash\hspace{0pt}}m{#1}}
\usepackage{tabularx} 
\newcolumntype{C}{>{\centering\arraybackslash}X} 
\begin{document}
% \newcommand{\aliasAPP}{CLUE}

% \title{\ourSystem: Safe Model-Based RL HVAC \underline{C}ontro\underline{L} Using Epistemic \underline{U}ncertainty \underline{E}stimation}

\title{A Safe and Data-efficient Model-based Reinforcement Learning System for HVAC Control}

\author{Xianzhong~Ding \orcidlink{0000-0001-6114-2801},~\IEEEmembership{}
        Zhiyu~An \orcidlink{0009-0000-1627-4194},~\IEEEmembership{}
        Arya~Rathee\orcidlink{0009-0008-6844-0463},~\IEEEmembership{}
        and~Wan~Du\orcidlink{0000-0002-2732-6954},~\IEEEmembership{Member,~IEEE}
        % <-this % stops a space
        
\thanks{A preliminary version of this work was published in the Proceedings of ACM BuildSys 2023 \cite{an2023clue}.\\
\indent Xianzhong Ding and Zhiyu An contributed equally to this work.\\
\indent Xianzhong Ding, Zhiyu An and Wan Du are with the Department of Computer Science and Engineering, University of California, Merced, Merced, CA 95343, USA (e-mail: xding5@ucmerced.edu; zan7@ucmerced.edu; wdu3@ucmerced.edu). (Corresponding author: Wan Du.)\\
\indent Arya Rathee is with the Department of Electrical Engineering, University of California, Santa Cruz, Santa Cruz, CA 95064, USA (e-mail: arathee@ucsc.edu). This work was accomplished when Arya Rathee was an undergraduate student at UC Merced and conducted an internship in Dr. Wan Du’s research group.}

}% <-this % stops a space

\maketitle

\begin{abstract}
Model-Based Reinforcement Learning (MBRL) has been widely studied for Heating, Ventilation, and Air Conditioning (HVAC) control in buildings. One of the critical challenges is the large amount of data required to effectively train neural networks for modeling building dynamics. This paper presents \textit{CLUE}, an MBRL system for HVAC control in buildings. \textit{CLUE} optimizes HVAC operations by integrating a Gaussian Process (GP) model to model building dynamics with uncertainty awareness. \textit{CLUE} utilizes GP to predict state transitions as Gaussian distributions, effectively capturing prediction uncertainty and enhancing decision-making under sparse data conditions. Our approach employs a meta-kernel learning technique to efficiently set GP kernel hyperparameters using domain knowledge from diverse buildings. This drastically reduces the data requirements typically associated with GP models in HVAC applications. Additionally, \textit{CLUE} incorporates these uncertainty estimates into a Model Predictive Path Integral (MPPI) algorithm, enabling the selection of safe, energy-efficient control actions. This uncertainty-aware control strategy evaluates and selects action trajectories based on their predicted impact on energy consumption and human comfort, optimizing operations even under uncertain conditions. Extensive simulations in a five-zone office building demonstrate that \textit{CLUE}  reduces the required training data from hundreds of days to just seven while maintaining robust control performance. It reduces comfort violations by an average of 12.07\% compared to existing MBRL methods, without compromising on energy efficiency. Our code and dataset are available at \color{blue}\url{https://github.com/ryeii/CLUE}\color{black}.

\end{abstract}


\section[Introduction]{Introduction}
The field of Heating, Ventilation, and Air Conditioning (HVAC) control has seen extensive research on Reinforcement Learning (RL) techniques~\cite{ding2019octopus, zhang2018practical, park2020hvaclearn}. RL offers adaptability by learning control policies through direct interactions with the environment~\cite{an2023clue}. The prevalent approach, Model-Free Reinforcement Learning (MFRL), obtains optimal HVAC control policies through trial-and-error interactions with real-world buildings. However, MFRL requires a substantial number of interactions to converge, with experiments indicating a need for 500,000 timesteps (equivalent to 5200 days) to achieve desirable control performance. Although simulated building models can accelerate the training process, they demand highly accurate calibration, posing significant challenges~\cite{zhang2018practical, ding2019octopus}.

Model-Based Reinforcement Learning (MBRL) offers a significant advantage in optimizing HVAC controls compared to traditional Model Predictive Control (MPC) and MFRL methods~\cite{moerland2023model}. Unlike MPC, which relies on an accurately formulated building thermal dynamics model—a challenging and often impractical task \cite{agbi2012parameter}, MBRL utilizes historical data to train a dynamics model and subsequently determine optimal control actions~\cite{ding2020mb2c, chen2019gnu}. On the other hand, MFRL, despite its direct policy learning approach through interaction with the building environment, suffers from extensive convergence times, often requiring years to yield practical results~\cite{ding2020mb2c,zhang2018practical}. A critical issue with MBRL-based HVAC control is the substantial volume of training data needed for model convergence. For instance, the $MB^2C$ (Multi-zone HVAC Control with Model-Based Deep
Reinforcement Learning) approach ~\cite{ding2020mb2c} models the building dynamics using an ensemble of deep neural networks combined with a Model Predictive Path Integral (MPPI) algorithm to optimize control actions. This Deep Ensemble (DE) model necessitates $183$ days of training data to achieve reliable predictions for effective control~\cite{ding2020mb2c, ding2024multi}.

Extensive experiments conducted on a simulated five-zone office building reveal that despite using years of training data, the model still exhibits epistemic uncertainty due to biases in the training data, resulting in inaccurate predictions for infrequently encountered states. To counteract this issue, we introduce an MBRL-based HVAC control approach designed to accommodate inaccuracies from models trained on limited datasets. This approach incorporates an awareness of the model’s prediction uncertainty, prompting conservative actions when predictions are uncertain. Enabling safe HVAC control necessitated an initial quantification of the epistemic uncertainty inherent in the building dynamics models.

Epistemic uncertainty estimation has significantly enhanced control performance in fields like robotic motion planning \cite{baek2023uncertainty} and reinforcement learning \cite{chua2018deep}. For scenarios with low-dimensional, discrete state spaces, such as multi-armed bandits, the count-based method \cite{vermorel2005multi, tang2017exploration} is often employed. This method quantifies model error by counting the frequency of training data instances corresponding to each input. However, this approach is less effective for building dynamics, which involve high-dimensional continuous variables where count-based methods struggle. Conversely, the DE method \cite{lakshminarayanan2017simple, chua2018deep} employs multiple neural networks to generate predictions, using the variance among these predictions to measure uncertainty. As introduced above, it requires a large amount of data to train the neural networks for HVAC control.

To bridge the existing gap in HVAC control, we introduce \textit{CLUE}, a novel MBRL system that leverages a Gaussian Process (GP) model for uncertainty-aware modeling of building dynamics. The GP model accepts inputs such as the state of the building (zone temperature, outdoor temperature, occupancy, etc.) and action (heating and cooling setpoints) and predicts the next state as a Gaussian distribution with both a mean and variance. The variance, indicative of the prediction's uncertainty, increases in scenarios with sparse data, serving as a measure of epistemic uncertainty. Research has shown that GP models can surpass conventional methods like neural networks, random forests, and support vector machines in minimizing model error \cite{MASSAGRAY2016119, goliatt2018modeling}.

The integration of GP models with MPPI for uncertainty-aware control represents a significant advancement in HVAC control. This approach allows for explicit modeling of prediction uncertainty, enhancing decision-making processes under uncertain conditions. By incorporating GP-derived uncertainty estimates, the MPPI algorithm can select safer and more energy-efficient actions, which is particularly important in environments where data is sparse or biased. However, integrating GP into a robust and efficient HVAC control system necessitates novel approaches within \textit{CLUE}: 1) Despite GP's non-parametric nature, it relies on a predefined kernel function with specific hyperparameters. We have developed a novel training method called meta kernel learning, which significantly enhances the efficiency of setting GP hyperparameters. 2) The GP-derived uncertainty is integrated into an MPPI algorithm, enabling the identification of safe, energy-efficient actions.

Selecting appropriate kernel parameters for GP with limited data is a challenge \cite{nghiem2017data}. Traditional methods for kernel selection typically require either expert experience to manually determine parameters or the use of gradient descent, which demands substantial data volumes to develop an effective kernel \cite{duvenaud2014automatic}. Inadequate kernel parameters can significantly degrade GP performance. To overcome these limitations, we developed a \emph{meta kernel learning} approach. This technique leverages the availability of extensive datasets from various buildings, recognizing that while obtaining large amounts of historical data for a specific target building may be challenging, similar data from other buildings are often readily accessible. Meta kernel learning uses this reference data to automatically learn an optimal kernel initialization, thereby minimizing the learning requirements without sacrificing model accuracy. This method contrasts with traditional approaches by significantly reducing the need for large volumes of training data, making it more suitable for practical applications where data collection can be expensive and time-consuming.

To translate uncertainty estimates into safe control actions effectively, we devised a confidence-based control algorithm that enables the safe and efficient application of MBRL even when model predictions are imprecise. Utilizing the MPPI control algorithm, our system generates an array of potential action trajectories using our building dynamics model. Each trajectory, a sequence predicting future actions, is assessed by MPPI to select the one that optimizes for both energy savings and minimal human comfort violations. Initially, the controller acts on the first step of this optimal trajectory. Incorporating GP-based uncertainty estimates, we developed a two-stage, uncertainty-aware HVAC control algorithm. The first stage employs a confidence threshold to exclude trajectories whose initial actions exhibit high uncertainty, preventing their execution to avoid potential risks. To establish the optimal confidence threshold for ensuring safe HVAC operations, we analyzed the relationship between the GP model's confidence values and the predicted model errors. This analysis helped in the development of an algorithm capable of optimizing this threshold offline by testing against historical data. The second stage involves refining the selection process using a modified MPPI objective function that considers the uncertainty of each action in all remaining trajectories. Additionally, we introduced a fallback mechanism, a reliable default control strategy activated when no safe actions are identified. This mechanism ensures the system maintains safe operation despite uncertainties.

We conducted extensive simulations to evaluate \textit{CLUE} in a $463m^2$ five-zone office building across three cities, using EnergyPlus \cite{doe2010energyplus}. Our initial tests focused on the accuracy of uncertainty estimation provided by the GP-based building dynamics model within \textit{CLUE}. The results showed that \textit{CLUE}'s building dynamics model significantly outperformed all baseline models in both modeling accuracy and uncertainty estimation. Subsequently, we applied \textit{CLUE} to manage HVAC controls within the simulated environment, comparing its performance in terms of human comfort and energy consumption against the state-of-the-art MBRL solutions. Our findings revealed that \textit{CLUE} consistently reduced comfort violations by an average of $12.07\%$ compared to the baseline MBRL methods. Additionally, \textit{CLUE} achieved comparable energy savings and showcased remarkable data efficiency by reducing the data requirement from hundreds of days to just seven days.

In summary, this paper brings several significant advancements to the field of HVAC control systems. Below are the four main contributions:

\begin{itemize}
    \item \textbf{Innovative MBRL System for HVAC Control:} Introduction of \textit{CLUE}, a novel MBRL system that utilizes a GP model for uncertainty-aware modeling of building dynamics. This system is tailored to enhance safety and efficiency by incorporating uncertainty directly into the control process. Our approach reduces the data requirements typically associated with GP models in HVAC applications by leveraging meta kernel learning, which optimizes GP hyperparameters using extensive datasets from various buildings, thus minimizing the need for large volumes of training data from a specific building.
    
    \item \textbf{Meta Kernel Learning:} Development of a new training method called meta kernel learning, which significantly improves the efficiency of setting GP hyperparameters. This method leverages extensive datasets from various buildings to reduce learning requirements without sacrificing model accuracy. This approach effectively reduces the data requirements by utilizing reference data from other buildings to enhance the GP model's performance, thereby addressing the challenge of sparse or noisy data in specific building scenarios.
    
    \item \textbf{Uncertainty-Aware Control Algorithm:} Design of a control algorithm that integrates GP-based uncertainty estimates with the MPPI algorithm. This includes mechanisms for excluding uncertain action trajectories and refining the selection process to ensure optimal and safe control actions. 
    
    \item \textbf{Extensive Simulation and Evaluation:} Comprehensive simulations were conducted in a five-zone office building across three cities to evaluate \textit{CLUE}'s performance. These simulations demonstrated that \textit{CLUE} significantly outperforms existing baselines in modeling accuracy, uncertainty estimation, human comfort, and energy efficiency. It also showcased remarkable data efficiency, reducing the data requirement from hundreds of days to just seven days. 
\end{itemize}

The remainder of this paper is organized as follows. Section \ref{sec: related-work} provides an overview of related work in HVAC control systems, focusing on various model-based and model-free reinforcement learning techniques. Section \ref{sec:mbrl} introduces the application of MBRL to HVAC control, covering state/action variables, optimization strategies, and a reward function balancing comfort and energy efficiency. Section \ref{sec: motivation} presents motivation experiments to understand model errors caused by data sparsity in state-of-the-art MBRL methods. Section \ref{sec: clue_design} details our proposed method, CLUE, including its architecture, modeling of building dynamics with GP, the meta-kernel learning technique, and the implementation of the confidence-based control mechanism. Section \ref{sec: evaluation} describes the experimental setup, including the simulation environment, performance metrics, and baseline comparisons. Section \ref{sec: discussion} discusses the limitations and future research directions. Finally, Section \ref{sec: conclusion} concludes the paper with a summary of our findings.

\section{Related Work}\label{sec: related-work}
There are a number of approaches to solving HVAC energy optimization problems in the literature, including model predictive control \cite{beltran2014optimal, li2023economic, winkler2020office}, Model-Free RL for HVAC \cite{park2020hvaclearn, zhang2018practical, vazquez2020marlisa, gao2020deepcomfort, ding2019octopus, lei2022practical, chen2020gnu}, Model-based RL for HVAC control \cite{zhang2019building, ding2020mb2c, chen2022mbrl}, Model-based approaches and Complementary Model-Free
Approaches. \cite{baldi2018automating, korkas2018grid, michailidis2015proactive, baldi2015model}, Safe RL for General and HVAC Applications \cite{an2023clue, an2024go, jayant2022model, liu2022safe, zhang2022safe}, and Adaptive-Predictive and Event-Based Control Strategies \cite{buonomano2015adaptive, short2012real, yang2019adaptive, schmelas2015adaptive, tesfay2018adaptive, tanaskovic2017robust, jia2018event, patti2014event, wu2015optimal, dhar2017adaptive, sun2015event, wang2016event, wang2018event}.

\noindent\textbf{MPC for HVAC control.} MPC is an iterative strategy used to solve optimal control problems, utilizing a receding time horizon to make decisions. In HVAC control, MPC has been instrumental in reducing energy consumption while maintaining occupant comfort. For example, Beltran et al. have developed an MPC framework tailored specifically for HVAC systems, focusing on optimizing energy efficiency and comfort \cite{beltran2014optimal}. More recently, the OFFICE framework has been introduced by Winkler et al., which balances energy costs and occupant comfort using a gray-box model. This model combines first principles with parameters that are dynamically updated, reflecting real-time system dynamics \cite{winkler2020office}. Further building on this concept, Li et al. proposed an economic model predictive control (EMPC) approach. This method optimizes energy consumption and indoor comfort by employing a lattice piecewise linear approximation of the Predicted Mean Vote (PMV) index, enhancing both the economic and comfort aspects of building management \cite{li2023economic}. However, the broader adoption of MPC in HVAC is hindered by significant challenges, primarily due to its dependency on precise models. Modeling is estimated to consume up to 75\% of the resources needed for MPC implementation, a notable barrier given the diverse nature of building environments \cite{rockett2017model}. Each building or thermal zone requires a custom model to effectively manage its unique characteristics, adding complexity and cost to the deployment of MPC systems \cite{killian2016ten, privara2013building, lu2015modeling}.

\noindent\textbf{Model-Free RL for HVAC.} Recent advancements in MFRL have demonstrated significant potential for optimizing HVAC controls. These techniques utilize learning agents that develop policies through active interaction with the environment and extensive trial and error. For example, traditional RL has been used to effectively adjust thermostat set-points, achieving a balance between energy efficiency and occupant comfort \cite{park2020hvaclearn}. DRL implementations have also been successful in real-life settings, managing radiant heating systems in office buildings \cite{zhang2018practical}. In tropical regions, Le et al. \cite{le2021deep} have deployed DRL to control air-free cooled data centers. Furthermore, Vazquez-Canteli et al. \cite{vazquez2020marlisa} developed a multi-agent RL strategy to optimize load shaping in grid-interactive buildings. Additionally, Gao et al. \cite{gao2020deepcomfort} used Deep Deterministic Policy Gradients (DDPGs) to formulate thermal comfort control policies, significantly enhancing HVAC system performance. Despite these successes, the primary focus of these studies remains on the HVAC subsystem. Extended research has introduced comprehensive building control frameworks that integrate HVAC, lighting, and window subsystems. These frameworks employ advanced algorithms, such as the Branching Dueling Q-network (BDQ), to improve control efficacy \cite{ding2019octopus, lei2022practical, ding2024exploring}. Nonetheless, RL applications in HVAC are challenged by high sample complexity, which necessitates prolonged training periods to develop robust control strategies, particularly in environments with large state-action spaces. In contrast, Gnu-RL utilizes a differentiable MPC policy that integrates domain knowledge, thereby enhancing both planning and system dynamics. This approach is noted for its data efficiency and interpretability \cite{chen2020gnu}. However, its reliance on the local linearization of water-based radiant heating system dynamics may restrict its applicability to more complex scenarios.

\noindent\textbf{Model-based RL for HVAC control.} Collecting large-scale real-world data for HVAC systems presents significant challenges. To address sample complexity, researchers have increasingly adopted MBRL techniques. Zhang et al. \cite{zhang2019building} developed an MBRL strategy where a neural network learns complex system dynamics, which are then integrated into an MPC framework using rolling horizon optimization for executing control actions. Similarly, Chen et al. \cite{chen2022mbrl} proposed MBRL-MC, a novel control strategy that combines MBRL with MPC. This strategy starts with the supervised learning of thermal dynamics for a specific zone and then designs a neural network planning framework that merges RL with MPC, diverging from traditional MBRL methods. It avoids the compounding error problem by not mimicking the outcomes of MPC's random shooting and excludes the bootstrapping technique from critical network updates, thereby enhancing stability. Despite their potential, MBRL methods typically excel in scenarios with lower state and action dimensions, such as in single-zone buildings. However, their performance often diminishes in more complex multi-zone buildings. Ding et al. \cite{ding2020mb2c} extended these capabilities to multi-zone buildings employing an ensemble of environment-conditioned neural networks dynamics models and a more efficient MPPI controller, achieving satisfactory results after 183 days of training data. 

\noindent\textbf{Model-based approaches and Complementary Model-Free Approaches.} Baldi et al. \cite{baldi2018automating} and Korkas et al. \cite{korkas2018grid} have explored model-based strategies for HVAC control, utilizing EnergyPlus for model design. Baldi et al. presented a switched self-tuning approach for automating occupant-building interaction via smart zoning of thermostatic loads, demonstrating significant improvements in energy efficiency and occupant comfort \cite{baldi2018automating}. Similarly, Korkas et al. proposed a distributed control and human-in-the-loop optimization for grid-connected microgrids, which has applications in HVAC systems \cite{korkas2018grid}. These works align with our approach by leveraging EnergyPlus for precise model-based control, highlighting both the benefits and challenges of using simulation-based models for HVAC optimization.

Complementary model-free approaches also offer promising solutions for HVAC control to reduce data requirements, particularly in complex and dynamic environments. Michailidis et al. \cite{michailidis2015proactive} demonstrate a proactive control strategy in a high-inertia building that uses simulation-assisted methods to optimize energy usage and thermal comfort, illustrating the effectiveness of model-free approaches in complex settings. Baldi et al. \cite{baldi2015model} introduced Parametrized Cognitive Adaptive Optimization (PCAO), designing efficient "plug-and-play" building optimization and control systems that learn optimal strategies with minimal human input, significantly improving energy efficiency and thermal comfort. These studies not only align with but also enrich our research objectives with innovative approaches to reduce data requirements and increase iteration efficiency. However, despite these advancements, building models often lack uncertainty awareness and require extensive training periods to achieve satisfactory performance.

\noindent\textbf{Safe RL for General and HVAC Applications.}  
Safe RL has become essential for ensuring the deployment of RL agents in real-world environments without causing harm \cite{ding2022drlic, an2023clue, an2024reward}. One prominent approach is Constrained Policy Optimization, which integrates safety constraints within the RL framework to prevent agents from taking harmful actions \cite{achiam2017constrained}. Although effective in theory, defining constraints that accurately reflect safety requirements is challenging. Model-Based Safe RL, another approach, uses simulated outcomes from learned environment models to enable safer exploration \cite{jayant2022model}. However, these models can sometimes yield suboptimal decisions due to inaccuracies, especially when applied to real-world scenarios. 

In the context of HVAC systems, ensuring safety during real-world exploration is critical, as outcomes of certain actions can be unpredictable. Research has suggested using simulators to train RL agents before deploying them in actual building settings \cite{zhang2018deep, zhang2018practical}. However, this method faces challenges due to potential mismatches between simulator predictions and real-world dynamics, which can cause "safe" decisions in simulations to become risky in practice. Alternatively, batch RL methods have been employed, utilizing historical data to refine and enhance control policies without direct interaction with the environment during training \cite{zhang2022safe, liu2022safe}. This method reduces risk but depends heavily on the data's quality and volume, influencing the effectiveness and safety of the control strategies implemented.

\noindent\textbf{Adaptive-Predictive and Event-Based Control Strategies.} Adaptive-predictive control strategies adjust control parameters in real-time based on environmental changes and system performance, enhancing the adaptability and responsiveness of HVAC systems. Event-based control strategies improve performance by responding to specific triggers or events rather than following a fixed schedule, offering potential for more dynamic and efficient HVAC management. Integrating these strategies with MBRL systems like CLUE could further optimize energy use and occupant comfort. Recent advancements in sensing technologies, communication networks, and embedded systems have enabled the development of sophisticated adaptive-predictive control strategies. These strategies use predictive models to anticipate environmental changes and adjust control parameters accordingly, providing more precise and responsive HVAC management \cite{buonomano2015adaptive, short2012real, yang2019adaptive, schmelas2015adaptive, tesfay2018adaptive, tanaskovic2017robust}. Event-based control strategies, on the other hand, focus on reducing unnecessary control actions by only responding to significant changes or events, thus improving energy efficiency and system performance \cite{jia2018event, patti2014event, wu2015optimal, dhar2017adaptive, sun2015event, wang2016event, wang2018event}. These methods have shown promise in managing the complexities of HVAC systems by dynamically adjusting to varying conditions, which is particularly beneficial in distributed and large-scale applications \cite{wu2015optimal, du2023optimizing}. The integration of adaptive-predictive and event-based control strategies with our proposed CLUE system offers a promising direction for future research, providing a framework that can leverage real-time data and event-driven triggers to enhance HVAC control performance.

\section{MBRL for HVAC Control}\label{sec:mbrl}
HVAC control can be modeled as a Markov Decision Process (MDP), represented as $\mathcal{M}:\{\mathcal{S}, \mathcal{A}, r, \mathcal{P}, \gamma\}$. This includes a state space $\mathcal{S}$, an action space $\mathcal{A}$, a reward function $r:\mathcal{S}\times\mathcal{A}\rightarrow\mathbb{R}$, a dynamics function $\mathcal{P}(s'|s, a)$, and a discount factor $\gamma$. In this framework, at each time step $t$, the controller is in state $s_t\in\mathcal{S}$, takes an action $a_t\in\mathcal{A}$, receives a reward $r_t=r(s_t, a_t)$, and moves to a new state $s_{t+1}$ determined by $s_{t+1}\sim\mathcal{P}(s_t, a_t)$. The objective at every step is to select an action that maximizes the discounted sum of future rewards, calculated as $\sum_{t'=t}^\infty\gamma^{t'-t}r(s_{t'}, a_{t'})$, where $\gamma\in[0, 1]$ weights the importance of immediate versus future rewards.

MBRL-based control integrates two core components: the dynamics model and the controller. The dynamics model, denoted by $f_\theta(s_t, a_t)$ and parameterized by $\theta$, leverages historical data $\{(s_t, a_t, s_{t+1})\}_n$ to predict the next state $s_{t+1}$ from the current state $s_t$ and action $a_t$. This prediction informs the controller's decision-making process, which involves solving the optimization problem:
\begin{equation}\label{Eq:background optimization problem}
    (a_t, \cdots, a_{t+H-1}) = \arg\max_{(a_t, \cdots, a_{t+H-1})}\sum_{t'=t}^{t+H-1}\gamma^{t'-t}r(s_{t'}, a_{t'})
\end{equation}
Here, the controller aims to select an action sequence that maximizes the cumulative discounted rewards over the next $H$ time steps. Typically, only the first action in this sequence is executed before re-evaluating and updating the plan at each subsequent time step based on the new state. While the random shooting method \cite{zhang2019building} has been traditionally used to address the optimization challenge in Eq.\ref{Eq:background optimization problem}, recent advancements indicate that the MPPI strategy offers superior optimization outcomes \cite{ding2020mb2c}. Our implementation of MBRL-based HVAC control utilizes insights from $MB^2C$ \cite{ding2020mb2c}, which outlines three components of MBRL.

% \subsubsection{State Design.}

\begin{center}
    \begin{table}[]
    \caption{State and action variables}
    \begin{center}
    \begin{tabular}{c|l}
    \hline
       Disturbances & Outdoor Air Drybulb Temperature ($^\circ C$) \\
       & Outdoor Air Relative Humidity (\%) \\
       & Site Wind Speed ($m/s$) \\
       & Site Total Radiation Rate Per Area ($W/m^2$) \\
       & Zone People Occupant Count ($No.$) \\
    \hline
       Zone State & Zone Air Temperature ($^\circ C$) \\
    \hline
       Action & Zone Temperature Setpoint ($^\circ C$)\\
    \hline
    \end{tabular}
    \end{center}
    \label{table_variables}
\end{table}
\end{center}

\textbf{States:} A state in the building dynamics model comprises a set of variables serving as both inputs and outputs. We categorize states into two groups: disturbances and the zone state, detailed in Table \ref{table_variables}. Disturbances include variables independent of HVAC system actions, such as weather conditions and occupancy levels. In contrast, the zone state refers specifically to the temperature of the controlled thermal zone, which is directly influenced by our control actions and is critical for calculating the reward of the building system.

\textbf{Actions:} The action within our system is defined as the temperature setpoint for the controlled thermal zone. On our experimental platform, the HVAC system's temperature setpoints range from a minimum of $15^\circ C$ to a maximum of $30^\circ C$. Each controlled thermal zone is equipped with both a heating and a cooling setpoint, leading to two action dimensions per zone.

\textbf{Rewards:}\label{Sec: reward design}
Our experimental platform utilizes the reward function outlined in \cite{jimenez2021sinergym}, as expressed by Eq. \ref{Eq: reward}. The comfort zone is defined between two temperature bounds, $\underbar{z}$ and $\overline{z}$, representing the lower and upper limits of comfortable zone temperature, respectively. At each timestep $t$, $Z_t$ indicates the current zone temperature, and $E_t$ quantifies the total energy consumption. To effectively balance comfort against energy usage, a weighting factor $w_e\in[0, 1]$ is employed. This factor adjusts the priority between maintaining comfort and minimizing energy consumption based on system needs.

\begin{equation} \label{Eq: reward}
    r(s_t) = - w_eE_t - (1-w_e)(|Z_t - \overline{z}|_+ + |Z_t - \underbar{z}|_+)
\end{equation}

In Eq. \ref{Eq: reward}, the weight $w_e$ is set to 0.1 during occupied periods to prioritize comfort, and to 1 during unoccupied periods to focus on energy savings. The energy term $E_t$ is derived using the L-1 norm of the difference between the action (setpoint temperature) and the actual zone temperature, representing the HVAC system’s heating or cooling effort \cite{chen2019gnu}. The comfort limits $\underbar{z}$ and $\overline{z}$ are set to $20^\circ C$ and $23.5^\circ C$ during winter, and $23^\circ C$ and $26^\circ C$ during summer, respectively.

\section{Motivation}\label{sec: motivation}
To evaluate the predictive performance of state-of-the-art MBRL methods \cite{zhang2019building, ding2020mb2c}, we conducted a series of EnergyPlus simulations \cite{doe2010energyplus, modes} within a 463 m$^2$ building comprising five zones \cite{jimenez2021sinergym, TODOS}. Utilizing actual 2021 TMY3 weather data from Pittsburgh, PA \cite{jimenez2021sinergym}, we initially focused on a single climate for preliminary experiments, later extending our analysis to encompass three distinct climate zones. Our investigation into the prediction errors of building dynamics models involved the application of the DE method detailed in \cite{ding2020mb2c}. Specifically, we examined three critical aspects: the correlation between model errors and the distribution of training data, the influence of varying training data sizes, and the temporal distribution of significant model errors.

\begin{figure*}
\begin{minipage}[t]{.35\linewidth}
    \includegraphics[width=\textwidth]{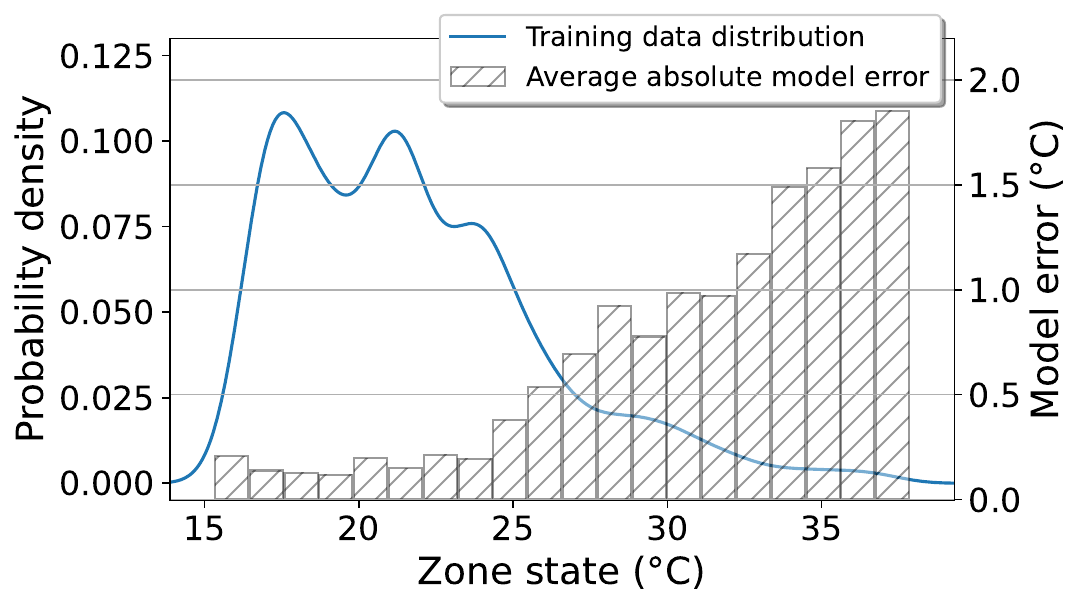}%
    \caption{Model errors are $>10\times$ higher in data-sparse regions vs. data-dense regions}%
    \label{fig:model_errors}    
  \end{minipage}\hfil
  \begin{minipage}[t]{.29\textwidth}
    \includegraphics[width=\linewidth]{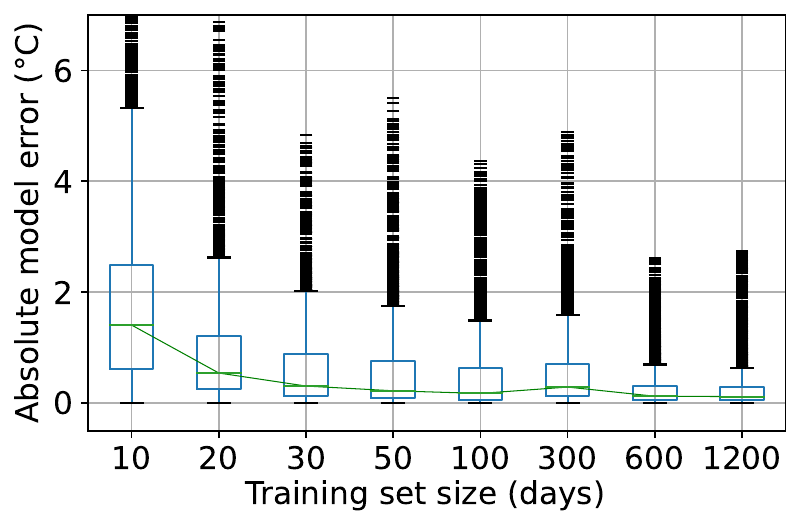}%
    \caption{Model error distribution vs. training data set size}%
    \label{fig:error_mot}
  \end{minipage}\hfil
  \begin{minipage}[t]{.31\textwidth}
    \includegraphics[width=\linewidth]{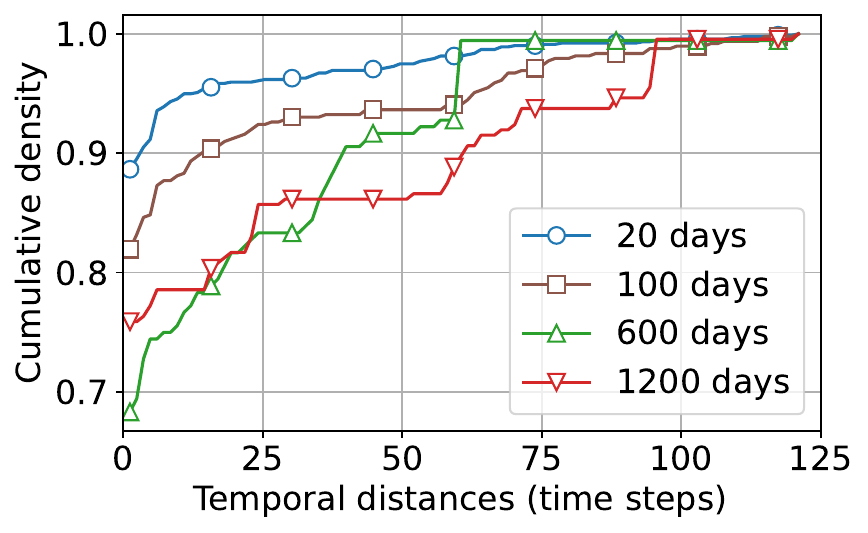}%
    \caption{CDF of the distances between model errors for each training set size}%
    \label{fig:error_distances}
  \end{minipage}%
\end{figure*}

\textbf{Experiment results:} Figure \ref{fig:model_errors} illustrates the relationship between the training data distribution and the prediction errors of the building dynamics model. The model was trained on $120,000$ time steps ($3.42$ years) using data collected via the default rule-based controller, across $150$ epochs with a learning rate of $1e-3$ to ensure convergence. After training, the model was tasked to predict the next $3000$ time steps, during which we recorded the prediction errors and the initial zone states for each prediction. To enhance the analysis, model errors were categorized based on the input zone temperatures, and average errors were computed for each bin.

In Figure \ref{fig:model_errors}, our analysis revealed that the DE model demonstrated significantly higher prediction errors in regions of the state space with sparse data. Particularly, this model exhibited enhanced precision when the zone state temperatures were between $15^\circ C$ and $25^\circ C$. While the average model error across all conditions was $0.29^\circ C$, we observed that errors consistently exceeded $1^\circ C$ in scenarios where the zone temperatures rose above $32^\circ C$. The highest model error recorded was $9.36\times$ greater than the average and remarkably $15,325 \times$ larger than the smallest model error.

\begin{figure}
\includegraphics[width=\columnwidth]{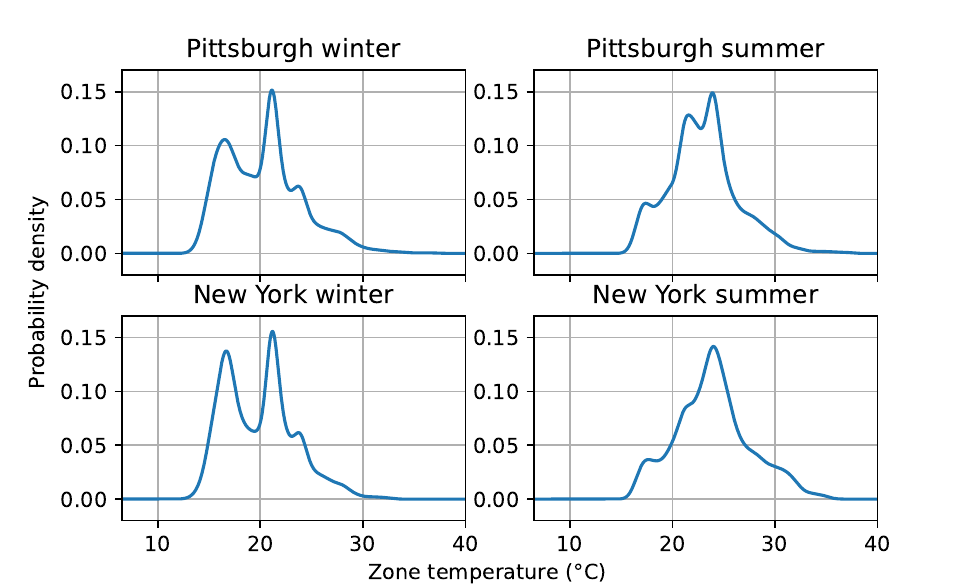}
\caption{KDE of the collected training data used for the dynamics model in MBRL-based HVAC control.}
\label{fig:pdf_of_collected_data}
\end{figure}

To understand the significant prediction errors observed in the DE models, we utilized Kernel Density Estimation (KDE) to examine the density distributions of the training data, which consisted of 10,000 transitions. The results are depicted in Figure \ref{fig:pdf_of_collected_data}. The blue curve illustrates the typical distribution patterns of the training data for the dynamics model in MBRL-based HVAC control. Notably, it features two distinct peaks around $17.5^\circ C$ and $21^\circ C$, corresponding to the typical night and day temperatures, respectively. During summer, these modes often merge into a single peak. The majority of the transition data tend to cluster around these central modes, which highlights an intrinsic bias in the thermal data. This bias significantly impacts the accuracy of predictions made by data-driven dynamics models.

% \begin{figure}
% \includegraphics[width=\columnwidth]{images/pdf_of_collected_data.pdf}
% \caption{PDFs of the collected data}
% \label{fig:pdf_of_collected_data}
% \end{figure}

% \begin{figure*}
% \includegraphics[width=\textwidth]{images/overview.pdf}
% \caption{Overview of the proposed \textit{CLUE} system}
% \label{fig:overview}
% \end{figure*}

\begin{figure*}
\includegraphics[width=\textwidth]{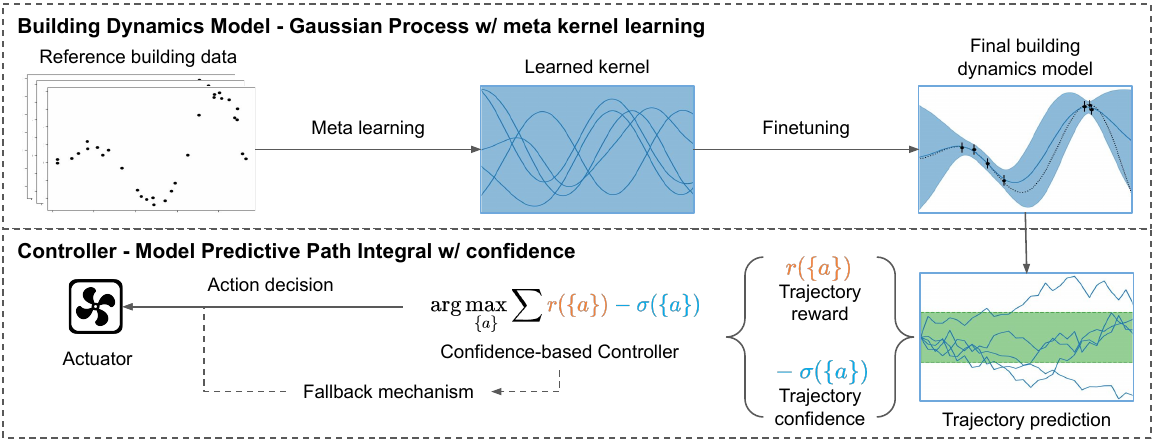}
\caption{Overview of the proposed \textit{CLUE} system. The framework integrates two key components: a dynamic model of building systems and a control mechanism based on MPPI control.}
\label{fig:overview}
\end{figure*}

\textbf{Two Questions Arise:} The first question centers on whether augmenting the size of historical data used for training models could reduce catastrophic model errors. Unfortunately, this approach does not yield the desired outcome. We trained a DE model using varied data sizes, from 10 to 1200 days, and evaluated its absolute error over the subsequent 30 days, as depicted in Figure \ref{fig:error_mot}. We deemed a model error exceeding 2°C as unacceptable because, in the building sensor domain, such a deviation is typically indicative of a sensor fault \cite{kumar2021building}. Despite utilizing nearly four years of data, errors surpassing 2°C were still evident; specifically, 2.1\% of all predictions deviated by more than 2°C from actual temperatures. This persistent error is largely due to epistemic uncertainty rooted in intrinsic biases within the thermal data of buildings, which significantly hampers the accuracy of predictions and leads to sub-optimal actions.

The second question explores whether temporal independence of model errors allows the thermal system to withstand brief controller discrepancies. Regrettably, this is not feasible. To investigate, we utilized the DE model to analyze each of the $3000$ time steps (representing one year) from previous experiments, recording model errors and specifically those exceeding $2^\circ C$. We reviewed these errors chronologically to determine the intervals between successive significant errors, plotting the cumulative density function of these intervals. This experiment was replicated with the DE model trained on datasets of varying sizes, from 20 to 1200 days, to observe changes related to training data volume. The results, depicted in Figure \ref{fig:error_distances}, indicate that a predominant majority (68\% to 89\%) of error intervals are zero, suggesting that substantial model errors tend to occur in sequences. This clustering of high errors, if overlooked, could result in prolonged periods of malfunction in the thermal control system.

\textbf{Our Key Idea:} Based on our observations, our primary objective is to mitigate the negative impacts of high model errors caused by distribution bias. Instead of refining the model or training process—which may be impractical—we have developed a procedure to preemptively alert the controller about potential high errors in the current time step. Essentially, our system \emph{anticipates} the uncertainty in the model's \emph{predictions} by analyzing its training data and the current input. When a high-error state-action pair is identified, we either disregard the prediction in favor of more reliable alternatives, or, if no prediction is deemed sufficiently reliable, we allow the building’s default controller to take over, thus addressing the lack of data in that region of the state space.

\section{The Design of \textit{CLUE}}\label{sec: clue_design}

In this section, we detail the architecture of \textit{CLUE}, focusing on problem formulation, dynamic modeling of the system using GP, and confidence-based MPPI control approach.

\subsection{Overview}
Figure \ref{fig:overview} illustrates the comprehensive framework of \textit{CLUE}. At a high level, \textit{CLUE} integrates two key components: a dynamic model of building systems and a control mechanism based on MPPI. Our dynamic model employs a GP to predict the future state of a building's HVAC system based on its current state, outputting both the predicted next state and a corresponding confidence interval. These predictions and their confidence intervals inform the MPPI controller's decisions, enabling it to select the most suitable actions.

The workflow of \textit{CLUE} is initiated by conducting meta-kernel learning (detailed in Section \ref{Sec: meta kernel learning}), which establishes the GP prior using reference data from other buildings or simulations, without prior knowledge of the target building $\mathcal{B}$. This learning phase helps adapt the GP kernel based on externally sourced data. Once sufficient data from the target building $\mathcal{B}$ is accumulated, the model fine-tunes the GP based on this building-specific data, incorporating the initial learned GP prior. This tailored GP model then serves as the dynamic model for the building.

The system sets an uncertainty threshold $\epsilon$ within the MPPI algorithm to manage decision-making risks. During deployment, \textit{CLUE} executes its confidence-based control strategy by generating multiple MBRL prediction trajectories using the GP model. It disregards any trajectory where the initial time step's uncertainty surpasses the threshold $\epsilon$. If all trajectories are discarded, a default action is dispatched to the building's actuator system. Conversely, if suitable trajectories remain, the MPPI leverages these to determine the optimal sequence of actions $a(\cdot)$, dispatching the first in this sequence to the actuator.

Following action deployment, \textit{CLUE} stops for a predefined interval (e.g., 15 minutes) to acquire new state data, which it then integrates into the historical dataset $\mathcal{D}_\mathcal{B}$, thereby readying the system for the next control cycle.

\subsection{Modeling Building Dynamics with GP} \label{Sec: Modeling System Dynamics with Gaussian Process}
We model the dynamics of the building using GP, interpreting the variance of each prediction as an indication of uncertainty—higher variance indicates higher uncertainty. The system's state, which includes environmental conditions (outdoor air temperature, humidity, occupancy, etc.) and zone-specific variables (zone air temperature), is consolidated into a state vector $s_t \in \mathcal{S}$. This state vector is then combined with the action vector $a_t \in \mathcal{A}$, which includes settings like heating and cooling setpoints, to form the input variable denoted as $x \in \mathcal{X} = \mathcal{S} \times \mathcal{A}$. The output variable is the subsequent zone state $s_{t+1}$. For simplicity and consistency, the input and output for each instance are denoted as $x_i$ and $y_i$, respectively, with the historical dataset defined as $\mathcal{D} = (X, Y) = \{(x_i,y_i)\}_{i=1}^n$.

Modeling with GP involves a two-stage process. It starts from a pool of candidate functions (GP prior) and computes beliefs conditioned on training data (GP posterior). We will introduce both stages in the following.

First, we select a function that intuitively quantifies the similarity between two inputs, denoted as $k: \mathcal{X}\times\mathcal{X} \rightarrow \mathbb{R}$. This function is the GP kernel. The idea is that buildings under similar conditions performing similar actions should transition to similar future states. Thus, if $k(x_1, x_3) > k(x_2, x_3)$, it suggests that $y_1$ is more relevant to predicting $y_3$ than $y_2$ is. Notably, each variable in our inputs contributes differently to the similarity to other inputs, depending on their level of impact according to the unknown dynamics function. The kernel function accommodates these differences using adjustable parameters, known as hyperparameters. \textit{CLUE} employs the Radial Basis Function (RBF) kernel, known for its expressive power in modeling complex systems such as building thermal dynamics, as represented in Eq. \ref{Eq: rbf kernel} and discussed in \cite{MASSAGRAY2016119}. The hyperparameters are $\theta: \{\theta_\text{scale}, \Theta\}$, where $\theta_{scale}$ is a scalar, and $\Theta$ is an $8\times8$ matrix. Overall, these parameters include a total of 65 real numbers. The selection of these parameters is critical as they significantly impact both the modeling accuracy and the uncertainty estimation accuracy.

The second stage involves fitting the data and making predictions. Using the exact regression technique, we first compute the covariance matrix $K := k(X, X)$, and then calculate the GP posterior using Eq. \ref{Eq:GP_post_mean} and Eq. \ref{Eq:GP_post_kernel}. Here, $I$ represents the identity matrix, and $\sigma_n^2$ is the noise variance (set to zero by default) to account for aleatoric uncertainty. From this posterior, the prediction of $y_*$ at $x_*$ follows a Gaussian distribution as specified in Eq. \ref{Eq:GP_pred}.

\begin{equation} \label{Eq: rbf kernel}
    k(x,x')=\theta_\text{scale}\exp\left(-\frac{1}{2}(x-x')^\top\Theta^{-2}(x-x')\right)
\end{equation}
\begin{equation} \label{Eq:GP_post_mean}
    m_\text{post}(x) = m(x)+k(x, X)(K+\sigma_n^2I)^{-1}(y-m(X))
\end{equation}
\begin{equation} \label{Eq:GP_post_kernel}
    k_\text{post}(x,x') = k(x, x')-k(x, X)(K+\sigma_n^2I)^{-1}k(X, x')
\end{equation}
\begin{equation} \label{Eq:GP_pred}
    \mathcal{GP}(x_*) = \mathcal{N}(m_\text{post}(x_*), k_\text{post}(x_*,x_*))
\end{equation}

The prediction mean $m_\text{post}(x_*)$ represents the most likely model outcome, while the variance $k_\text{post}(x_*,x_*)$ indicates the level of uncertainty associated with the model outcome given data $\mathcal{D}$ and input $x_*$. This variance, serving as the indicator of epistemic uncertainty, is higher if $x_*$ is in a data-scarce region of the input space $\mathcal{X}$, emphasizing areas of greater uncertainty and guiding decision-making under these conditions.

To clearly understand the GP model integration process, we outline the steps involved:

\textbf{Steps for GP Model Integration:}
\begin{enumerate}
    \item \textbf{Input Representation:} Consolidate state and action variables into input vectors $x_t \in \mathcal{X} = \mathcal{S} \times \mathcal{A}$, where $\mathcal{S}$ represents state space and $\mathcal{A}$ represents action space.
    \item \textbf{Kernel Function Selection:} Employ the RBF kernel for its expressive power in modeling complex building dynamics.
    \item \textbf{GP Training:} Use historical data $\mathcal{D} = \{(x_i, y_i)\}_{i=1}^n$ to train the GP model, optimizing kernel hyperparameters through meta-kernel learning.
    \item \textbf{Prediction and Uncertainty Estimation:} For a given input $x_*$, predict the next state as a Gaussian distribution $\mathcal{GP}(x_*) = \mathcal{N}(m_\text{post}(x_*), k_\text{post}(x_*, x_*))$, where $m_\text{post}(x_*)$ is the mean and $k_\text{post}(x_*, x_*)$ is the variance indicating prediction uncertainty.
\end{enumerate}

The steps for GP model integration provide a structured approach to implementing the GP model within \textit{CLUE}. The next section will discuss the meta-kernel learning technique, which is essential for optimizing the GP model's hyperparameters to enhance prediction accuracy and uncertainty estimation.

\subsection{Meta Kernel Learning} \label{Sec: meta kernel learning}
To model building dynamics with GP effectively, we optimize the kernel parameters' initialization using data from a varied collection of reference buildings. For this purpose, we have developed a novel meta kernel learning method that integrates meta-learning \cite{finn2017model} with kernel learning \cite{duvenaud2014automatic}.

With a given set of training data $\mathcal{D}:\{X, Y\}$, kernel learning optimizes the kernel parameters through gradient descent. It defines the loss function as the mean squared error (MSE) between the GP model's predictions and the actual data, $\mathcal{L}(\mathcal{GP}_{\theta_k}) = MSE(\mathcal{GP}_{\theta_k}(X), Y)$. The process involves calculating the gradient of the loss with respect to the kernel parameters, $\nabla_{\theta_k}\mathcal{L}(\mathcal{GP}_{\theta_k})$, and updating the parameters accordingly. This procedure repeats for $n$ iterations until a set of kernel parameters with minimized model error is obtained.

Meta-learning \cite{finn2017model} enhances an agent's ability to learn from a variety of tasks, allowing for quick adaptation to new tasks. This method moves beyond the standard optimization of model parameters, traditionally expressed as:

\begin{equation}
    \theta_k^* = \arg\min_{\theta_k}\mathcal{L}(\mathcal{GP}_{\theta_k}),
\end{equation}

Instead, meta-learning focuses on fine-tuning the parameters to reduce the aggregated loss across multiple tasks, which can be formalized as:

\begin{equation}
    \theta_k^* = \arg\min_{\theta_k}\sum\mathcal{L}_{\mathcal{T}_i}(\mathcal{GP}_{\theta_k})
\end{equation}

Here, $\mathcal{T}_i$ represents an individual task, and the aim is to minimize the overall model error by optimizing across a diverse set of tasks.

Our meta kernel learning process fine-tunes the GP kernel hyperparameters, $\theta_k$, through a tailored kernel learning approach. We redefine tasks typically associated with MDPs to focus on the modeling of building data over specific periods. For example, the task might involve optimizing the GP model's accuracy using building data from Pittsburgh over the summer months, encapsulated by the model loss function in Eq.~\ref{Eq:GP_Meta_loss}. In this equation, $X_{\mathcal{T}_i}$ represents the input features (e.g., historical temperature readings, occupancy levels) and $Y_{\mathcal{T}_i}$ denotes the corresponding output targets (e.g., future temperature readings) for task $\mathcal{T}_i$. This equation quantifies the discrepancy between GP predictions and actual observations when applying the optimized kernel hyperparameters, $\theta_k$.

The essence of our meta-learning approach is to minimize this model error aggregated over an entire suite of such tasks, collectively denoted by $p(\mathcal{T})$. The main objective function, aiming at comprehensive error reduction, is outlined in Eq. \ref{Eq:GP_Meta_learning}. By executing this strategy, we systematically enhance the kernel's initial parameter set, enabling more precise subsequent adaptations to the specific thermal dynamics of different buildings.

\begin{equation}\label{Eq:GP_Meta_loss}
    \mathcal{L}_{\mathcal{T}_i}(\mathcal{GP}_{\theta_k}) = MSE(\mathcal{GP}_{\theta_k}(X_{\mathcal{T}_i}), Y_{\mathcal{T}_i})
\end{equation}
% \vspace{-0.1in}
\begin{equation} \label{Eq:GP_Meta_learning}
    \theta_k^* = \arg\min_\theta\sum_{i=1}^n\mathcal{L}_{\mathcal{T}_i}(\mathcal{GP}_{\theta_k})|_{\mathcal{T}_i\sim p(\mathcal{T})}
\end{equation}

To clearly understand the meta-kernel learning process, we outline the steps involved:

\textbf{Steps for Meta-Kernel Learning:}
\begin{enumerate}
    \item \textbf{Data Collection:} Gather extensive datasets from various reference buildings, capturing diverse thermal dynamics.
    \item \textbf{Meta-Learning Initialization:} Randomly initialize the GP kernel parameters $\theta_k$. Utilize these reference datasets to further refine the initial parameters. Define tasks $\mathcal{T}_i$ for different building data subsets.
    \item \textbf{Kernel Parameter Optimization:} For each task, compute the loss $\mathcal{L}(\mathcal{GP}_{\theta_k}) = MSE(\mathcal{GP}_{\theta_k}(X_{\mathcal{T}_i}), Y_{\mathcal{T}_i})$ and update kernel parameters through gradient descent.
    \item \textbf{Meta-Optimization:} Aggregate losses across tasks and perform meta-optimization to refine kernel parameters, minimizing the overall model error.
\end{enumerate}

% \begin{algorithm}
% \caption{Meta Kernel Learning}
% \label{alg:meta}
% \SetKwProg{control}{Function \emph{control}}{}{end}
% \KwIn{$p(\mathcal{T})$ distribution over all building data, $\{\alpha, \beta\}$ step size hyperparameters}
% Randomly initialize $\theta_{k}$\\
% \While{not converged}{
%     Sample batch of building data $\{\mathcal{T}_1, \cdots, \mathcal{T}_i\}\sim p(\mathcal{T})$\\
%     \For{all $\mathcal{T}_i$}{
%         $\mathcal{GP}\gets\text{GP Fit}(X_{\mathcal{T}_i}, Y_{\mathcal{T}_i})$\\
%         $\theta'_k\gets\theta_k-\alpha\nabla_{\theta_k}\mathcal{L}_{\mathcal{T}_i}(\mathcal{GP}_{\theta_k})$\\
%     }
%     $\theta_k \gets\theta_k-\beta\nabla_{\theta}\sum\mathcal{L}_{\mathcal{T}_i}(\mathcal{GP}_{\theta_k'})
% }
% \end{algorithm}

\begin{algorithm}
\caption{Meta Kernel Learning}
\label{alg:meta}
\SetKwProg{control}{Function \emph{control}}{}{end}
\KwIn{$p(\mathcal{T})$ distribution over all building data, $\{\alpha, \beta\}$ step size hyperparameters}
Randomly initialize $\theta_{k}$\\
\While{not converged}{
    Sample batch of building data $\{\mathcal{T}_1, \cdots, \mathcal{T}_i\}\sim p(\mathcal{T})$\\
    \For{all $\mathcal{T}_i$}{
        $\mathcal{GP}\gets\text{GP Fit}(X_{\mathcal{T}_i}, Y_{\mathcal{T}_i})$\\
        $\theta'_k\gets\theta_k-\alpha\nabla_{\theta_k}\mathcal{L}_{\mathcal{T}_i}(\mathcal{GP}_{\theta_k})$\\
    }
    $\theta_k \gets\theta_k-\beta\nabla_{\theta}\sum\mathcal{L}_{\mathcal{T}_i}(\mathcal{GP}_{\theta_k'})$\\
}
\end{algorithm}

Our meta kernel learning approach is detailed in Algorithm \ref{alg:meta}, where kernel parameters are refined using gradients from multiple tasks, thus enhancing the kernel's general applicability and avoiding overfitting to a single data set. For the step sizes of hyperparameters, we have chosen $\alpha=\beta=1e-3$. This conservative step size slows the training pace but is essential for ensuring convergence in our experiments. The resulting set of kernel parameters forms the initial configuration for the GP model, which is then fine-tuned on target building data. After applying meta-learning, we fine-tune the pre-trained kernel with a small data subset from the target building, using standard kernel learning methods to achieve a tailored building-specific kernel.

The integration of meta-kernel learning into the GP model significantly enhances its performance by optimizing the kernel parameters using diverse datasets. The next section will discuss the confidence-based control approach, which utilizes these GP models to make informed control decisions.

\subsection{Confidence-based Control} \label{Sec: confidence-based control}

\textit{CLUE} employs online planning through MPPI control to determine actions. Starting from the current building state $s_t$ at time $t$, and considering a prediction horizon $H$, \textit{CLUE} utilizes the GP-based dynamics model $\mathcal{GP}(x_*)$ to predict the future states $s_{t:t+H}$ based on the proposed action sequence $a_{t:t+H} = \{a_t, \ldots, a_{t+H}\}$. At each timestep $t$, the MPPI controller implements the initial action $a_t$ from this sequence. The accuracy of the first predicted state in the sequence is crucial, as it greatly influences the overall performance of the control system. To ensure reliability, the model's uncertainty estimation, which does not always align directly with actual model error, is converted into a tangible model error threshold. This conversion is performed by evaluating the model against historical data. Subsequently, \textit{CLUE} screens and excludes any prediction trajectories where the uncertainty exceeds this predefined threshold, ensuring that only the most reliable trajectories influence control decisions.

To better understand how \textit{CLUE} implements this process, we outline the steps for the MPPI control approach:

\textbf{Steps for MPPI:}
\begin{enumerate}
    \item \textbf{Trajectory Generation:} Generate multiple action trajectories from the current state $s_t$, each predicting future states $s_{t:t+H}$ based on the GP dynamics model.
    \item \textbf{Uncertainty Filtering:} Exclude trajectories where the initial time step's uncertainty exceeds a predefined threshold $\epsilon$, ensuring only reliable trajectories are considered.
    \item \textbf{Trajectory Evaluation:} For each remaining trajectory, compute the objective function $\sum_{t=1}^H\gamma^t(r(x_t)-\lambda \sigma(x_t))$, balancing rewards and uncertainties.
    \item \textbf{Action Selection:} Select the trajectory with the highest objective value and execute the first action $a_t$.
\end{enumerate}

\subsubsection{Uncertainty Threshold Translation} \label{Sec: Uncertainty Threshold Translation}
% In our GP dynamics model, uncertainty at a given input $x$ is denoted by $\sigma(x)$. We define a model error threshold $e^*$ (in degrees Celsius) and an uncertainty flagging threshold $\epsilon \in \mathbb{R}^+$. If $\sigma(x) > $\epsilon$, the state-action pair is flagged as having high model error. The goal of our threshold translator is to optimize the identification of true model errors greater than $e^*$, effectively maximizing the detection of true positives and true negatives.

In our GP dynamics model, uncertainty at a given input $x$ is denoted by $\sigma(x)$. We define a model error threshold $e^*$ (in degrees Celsius) and an uncertainty flagging threshold $\epsilon \in \mathbb{R}^+$. If $\sigma(x) > \epsilon$, the state-action pair is flagged as having high model error. The goal of our threshold translator is to optimize the identification of true model errors greater than $e^*$, effectively maximizing the detection of true positives and true negatives.

To determine the optimal $\epsilon$, we utilize an offline procedure that leverages historical data from the target building, $\mathcal{D}:\{X, Y\}$. The GP model, $\mathcal{GP}$, predicts each input $x_i \in X$, yielding results $(\vec{\mu}, \vec{\sigma}) = \mathcal{GP}(X)$, where $\vec{\mu}$ represents predicted states and $\vec{\sigma}$ signifies associated uncertainties. The absolute model errors $\vec{e} = |Y - \vec{\mu}|$ are calculated, testing the model's accuracy against historical data. For all prediction pairs $\{(\mu, \sigma)\}_n$, we solve the following optimization problem:

\vspace{-1em}
\begin{equation}\label{Eq: Uncertainty Threshold Translation}
% \vspace{-0.5in}
\begin{split}
    \text{minimize: }& \text{count}(|y_i - \mu_i| < e^*)-\text{count}(|y_i - \mu_i| > e^*)\\
    \text{s.t. }& \sigma_{i} > \epsilon
\end{split}
% \vspace{-0.5in}
\end{equation}

% By setting $\epsilon$, we aim to maximize the model's accuracy in classifying errors larger than $e^*$, ensuring that $\sigma > $\epsilon$ accurately reflects high model errors. This objective aligns with our goal to establish the most effective uncertainty threshold $\epsilon$, which will be employed in subsequent predictions, using prediction accuracy as a surrogate measure to evaluate the optimization outcome.

By setting $\epsilon$, we aim to maximize the model's accuracy in classifying errors larger than $e^*$, ensuring that $\sigma > \epsilon$ accurately reflects high model errors. This objective aligns with our goal to establish the most effective uncertainty threshold $\epsilon$, which will be employed in subsequent predictions, using prediction accuracy as a surrogate measure to evaluate the optimization outcome.

To provide HVAC engineers and building managers with greater control, we introduce the following mechanism:

\textbf{Uncertainty threshold control knob.}  Our method accommodates any model error threshold, $e^*\in\mathbb{R}^+$, as specified by HVAC engineers. Each expected model error threshold is systematically converted into an uncertainty threshold. \textit{CLUE} performs this translation offline to determine the optimal flagging threshold, which is then utilized in the confidence-aware MPPI control process. This introduces an uncertainty threshold control knob, enabling HVAC engineers or building managers to set the maximum acceptable model error. Consequently, a lower threshold leads \textit{CLUE} to adopt more conservative actions, resulting in fewer violations but possibly higher energy usage. Conversely, a higher threshold encourages \textit{CLUE} to take bolder actions to conserve energy, which may increase the rate of violations. The impact of adjusting the uncertainty threshold control knob will be empirically evaluated in our upcoming experiments, as detailed in Section \ref{Sec: Effect of confidence threshold control knob.}.

\subsubsection{Confidence-Aware MPPI} \label{Confidence-Aware MPPI}
During each planning cycle, the MPPI controller generates multiple trajectories. \textit{CLUE} evaluates the uncertainty of the initial time step for each trajectory, comparing it against a predefined model error threshold established by the uncertainty threshold translator. Trajectories that exceed this threshold are flagged and subsequently discarded. This process ensures that only trajectories with acceptable uncertainty levels are considered for action selection, effectively removing those with high uncertainty from decision-making.

After filtering trajectories based on their initial uncertainties, the confidence-based MPPI selects the optimal trajectory from those that remain. While these trajectories have low initial uncertainties, it is crucial to consider uncertainties at future time steps as they impact the accuracy of predicted future rewards. \textit{CLUE} integrates this aspect into the optimization of the objective function, detailed in Eq. \ref{Eq: new mpc objective}, by using $\lambda$ to balance the trade-offs between uncertainty and reward. Specifically, Eq. \ref{Eq: new mpc objective} is designed to maximize the sum of discounted rewards while minimizing the sum of discounted uncertainties across the prediction horizon. This approach recognizes that as the controller rolls out trajectories step by step, the significance of initial model accuracy gradually decreases. Therefore, each uncertainty value is discounted by a rate $\gamma$, reflecting its diminishing influence over time. This adjustment to the objective function allows the controller to optimize simultaneously for high rewards and low uncertainties, enhancing decision-making under uncertainty.

\begin{equation} \label{Eq: new mpc objective}
    a(\cdot)^* = \arg\max_{a(\cdot)}\sum_{t=1}^H\gamma^t(r(x_t)-\lambda \sigma(x_t))
\end{equation}
\begin{equation} \label{Eq: MPPI}
    a_{\text{new}} = a_\text{prev}+\frac{\sum_{k=1}^K \delta a_k\exp(\frac{1}{\eta}\sum_{t=1}^H \gamma^t (r(x_t)-\lambda \sigma(x_t)))}{\sum_{k=1}^K \exp(\frac{1}{\eta}\sum_{t=1}^H \gamma^t (r(x_t)-\lambda \sigma(x_t)))}
\end{equation}

Integrating the newly designed optimization objective function into MPPI controllers is both straightforward and broadly applicable. For example, we demonstrate this process with the MPPI controller used in recent HVAC control research \cite{ding2020mb2c}. MPPI determines the optimal action sequence by evaluating expected rewards across randomly generated trajectories, then calculates the weighted sum of these trajectories’ action sequences using exponential weighting based on the cumulative discounted rewards \cite{williams2016aggressive}.

A standard MPPI implementation for HVAC control might use a reward function as described in Eq. \ref{Eq: reward}, where $R = \sum_{t=1}^H \gamma^t r(s_t)$, and $H$ is the length of the prediction horizon. To adapt our optimization method, we modify this reward function to align with the objective outlined in Eq. \ref{Eq: new mpc objective}. The adaptation results in Eq. \ref{Eq: MPPI}, where $a$ represents an action sequence, $K$ the number of trajectories, $\delta_{a_k}$ the action perturbation, and $\eta$ a hyperparameter. Eq. \ref{Eq: MPPI} illustrates how the weighted exponential sum of action sequences is now calculated concerning both the discounted rewards and their associated uncertainties. This method of reward evaluation can be similarly adopted by other MBRL controllers by substituting their reward functions with our proposed formulation.

\subsubsection{Fallback Mechanism} \label{Sec: fallback mechanism}
In situations where all trajectories generated by \textit{CLUE} exhibit high uncertainty, the system defaults to a rule-based controller. This default controller, commonly used in current HVAC systems, offers conservative and reliable actions, ensuring safety and dependability when predictive confidence is low.

\section{Evaluation}\label{sec: evaluation}
We conducted two comprehensive sets of experiments to thoroughly evaluate \textit{CLUE}, focusing on different aspects of its performance. The first set of experiments assessed the modeling accuracy of GP equipped with meta-kernel learning. This stage also tested the reliability of the uncertainty prediction mechanisms provided by the fallback strategy, critical for managing unpredictable scenarios. The second set of experiments involved deploying \textit{CLUE} within various simulated environments, where its performance was rigorously compared against current state-of-the-art solutions to determine its effectiveness in realistic settings.

\subsection{Experiment Setting}
\subsubsection{Platform Setup}
Our experimental platform was carefully chosen to ensure high fidelity and effective integration of software tools. We employed EnergyPlus \cite{doe2010energyplus}, a sophisticated simulation tool widely recognized for its accuracy in modeling building energy systems. For the deep learning components, PyTorch \cite{paszke2019pytorch} was selected for its flexibility and robustness in handling complex neural network architectures. We utilized GPyTorch \cite{gardner2018gpytorch}, an extension to PyTorch that provides advanced capabilities for GP computations, particularly for meta-kernel learning with GPU acceleration. This setup allowed for rapid iterations and real-time processing. Sinergym \cite{jimenez2021sinergym}, serving as the virtual testbed, enabled seamless interaction between \textit{CLUE} and the EnergyPlus environment. It managed the exchange of actions and state observations effectively, ensuring that the decision-making process of \textit{CLUE} was both reactive and informed based on real-time data feedback. All components used in our experiments are open-source, which promotes transparency and allows other researchers to replicate our results easily or extend them in future studies.

\subsubsection{Implementation details}
Throughout our experiments, we maintained consistent hyperparameters to ensure the reliability of our results. For DE models, we set the epochs to 150, the learning rate to \(1 \times 10^{-3}\), and the weight decay to \(1 \times 10^{-5}\). For GP kernel learning, we set the number of iterations to 800 with a learning rate of \(1 \times 10^{-2}\), and for fine-tuning the GP kernels, we used 200 iterations with the same learning rate. MSE served as the loss criterion, and we used the Adam optimizer for all training processes. In the context of meta-kernel learning, the GP model was trained on data spanning \(35,040\) time steps, equivalent to one year, from each reference building. For the MPPI controller, we adhered to the optimal hyperparameter configuration previously tested, with a sample number of 1000 and a horizon of 20 steps \cite{ding2020mb2c}. For the confidence-based MPPI, we set \(\lambda\) to \(1 \times 10^{-2}\).

\subsubsection{Environment selection}
Our simulation was conducted on a building with an area of \(463 \, m^2\) and five zones, as outlined in Sinergym \cite{jimenez2021sinergym}. We selected three climate-distinct cities in the United States—Pittsburgh, Tucson, and New York—for the experiments, which took place from January 1st to January 31st and July 1st to July 31st. Each city was chosen to represent a different climate type, according to ASHRAE standards: Pittsburgh features a continental climate (ASHRAE 4A), Tucson a hot desert climate (ASHRAE 2B), and New York a humid continental climate (ASHRAE 4A) \cite{standard2020ansi}. We used actual 2021 TMY3 weather data for these cities \cite{jimenez2021sinergym}. This selection strategy was intended to ensure the generalizability of \textit{CLUE} across diverse climate conditions, providing a robust test of its applicability in varied regional settings.

To ensure a comprehensive evaluation, we adopted specific parameters and evaluation metrics tailored to capture different aspects of \textit{CLUE}'s performance.

\subsubsection{Performance Metrics}
We use two different sets of performance metrics to evaluate the uncertainty estimation accuracy and building control efficiency respectively.

\textbf{1) Metrics for the uncertainty estimation:}
\begin{itemize}
 \item Accuracy: the sum of true positive and true negative results divided by the total number of results. Higher accuracy indicates that the model makes fewer mistakes in total.
 \item Precision: the number of true positive results divided by the number of all positive results, including those not identified correctly. Higher precision indicates that the model is more efficient, i.e. the model wastes less control steps from falsely flagging low model error predictions.
 \item Recall: the number of true positive results divided by the number of all samples that should have been identified as positive. Higher recall indicates that the model is safer, i.e. it correctly flags a larger portion of all high model error predictions.
\end{itemize}

Note that we did not use the $F_1$ score metric due to its equal weighting of precision and recall \cite{hand2018note}, which does not align with our control system's objective. Instead, we directly logged the precision and recall values for informative data representation.

\textbf{2) Metrics for building control efficiency:}
\begin{itemize}
 \item Cumulative reward: the weighted sum of the building system rewards calculated according to Eq. \ref{Eq: reward}. Higher cumulative reward indicates overall better constraint compliance and energy efficiency.
 \item Violation rate:  the ratio of time steps where zone temperature violates comfort constraints to the total number of time steps. A lower violation rate indicates improved constraint compliance and safer operation of the HVAC system.
 \item Energy consumption: total energy consumption in kWh. With a low violation rate, lower total energy consumption indicates higher energy efficiency.
\end{itemize}

\subsubsection{Baselines}
To evaluate our method's performance, we benchmarked it against several baselines in two key areas: uncertainty estimation accuracy and building control efficiency.
\paragraph{Uncertainty Estimation Accuracy}
We compared our method's uncertainty estimation capabilities with the following models:
\begin{itemize}
 \item \textbf{DE}: Utilizes an ensemble of five neural networks. Although the original study \cite{ding2020mb2c} does not focus on uncertainty estimation, we adopted the uncertainty measurement method described by Lakshminarayanan et al. \cite{lakshminarayanan2017simple}. Here, uncertainty is quantified as the variance of the predictions from the ensemble, calculated using the formula:
 \[
 \sigma_*^2(x) = \frac{1}{M}\sum_{m=1}^M \mu_{\theta_m}^2(x) - \mu_*^2(x),
 \]
 where \( M=5 \) represents the number of models, \( \mu_{\theta_m}(x) \) denotes the prediction of model \( m \), and \( \mu_*(x) \) is the average prediction across all models.
 \item \textbf{GP}: Employs a GP method with kernel learning, using an uncertainty estimation approach identical to ours. This represents a standard method for regression in building thermal dynamics.
\end{itemize}

\paragraph{Building Control Efficiency}
For evaluating building control efficiency, we benchmarked against the following systems:
\begin{itemize}
 \item \textbf{Rule-based}: This is the default controller provided by the simulation environment, serving as a basic comparison point.
 \item \textbf{DE-MBRL} \cite{ding2020mb2c}: An MBRL system that integrates a DE with MPPI control.
 \item \textbf{\textit{CLUE} w/o Confidence-Based control (CB)}: Our system, \textit{CLUE}, excluding the confidence-based control feature, to evaluate the impact of this component on overall system performance.
\end{itemize}

The comparison criteria focused on evaluating the precision, recall, and accuracy of uncertainty estimation as well as the cumulative reward, violation rate, and energy consumption for control efficiency. This comprehensive evaluation ensured a holistic understanding of \textit{CLUE}'s performance.

\begin{figure}
    \includegraphics[width=\columnwidth]{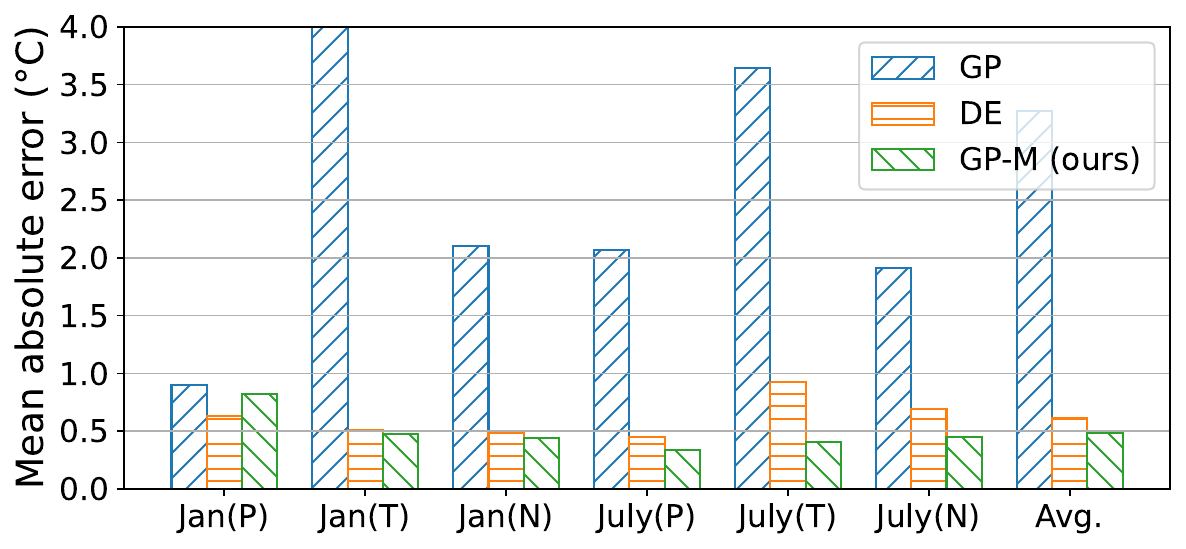}
    \caption{Model accuracy results comparing GP-M, DE, and standard GP.}
    \label{fig:compare_model_errors}
\end{figure}

\subsection{Modeling and Uncertainty Estimation}
\subsubsection{Modeling Accuracy}
We evaluated the performance of our solution, GP with meta-kernel learning (denoted as GP-M in Figure \ref{fig:compare_model_errors}), against two baseline methods: the DE and a standard GP. The DE model was trained using 2,000 time steps, equivalent to 20.83 days of data from the target building. Similarly, the GP model was trained and fitted using the same dataset. The GP-M approach commenced with meta-kernel learning before being fine-tuned and fitted to the same data as the DE model. Our findings reveal that GP-M achieved a mean absolute model error that was 20.7\% lower than that of the DE model and 85.1\% less than that of the standard GP model. In all tested environments, GP-M consistently outperformed the GP model and surpassed the DE model in five of the six environments. These results underscore the effectiveness of meta-kernel learning in providing a robust initialization for kernel parameters, thereby significantly enhancing the modeling performance of the GP approach.

\begin{table*}
\caption{Uncertainty estimation experiment results (best performances in each comparison are in \textbf{bold})}
\begin{center}
\begin{tabular}{|cc|cc|cc|cc|}
\hline
\multicolumn{2}{|c|}{Location}                                                          & \multicolumn{2}{c|}{Pittsburgh, PA}           & \multicolumn{2}{c|}{Tucson, AZ}               & \multicolumn{2}{c|}{New York, NY}             \\ \hline
\multicolumn{2}{|c|}{Time}                                                              & January               & July                  & January               & July                  & January               & July                  \\ \hline
\multicolumn{1}{|c|}{\multirow{3}{*}{DE \cite{lakshminarayanan2017simple}}} & Accuracy  & .796$\pm$.00          & .831$\pm$.01          & .736$\pm$.01          & .650$\pm$.00          & .830$\pm$.00          & .679$\pm$.00          \\
\multicolumn{1}{|c|}{}                                                      & Precision & .521$\pm$.01          & \textbf{.851$\pm$.09} & .439$\pm$.08          & .489$\pm$.00          & .403$\pm$.02          & .373$\pm$.01          \\
\multicolumn{1}{|c|}{}                                                      & Recall    & .740$\pm$.01          & .160$\pm$.10          & .693$\pm$.12          & .827$\pm$.00          & .816$\pm$.00          & .812$\pm$.01          \\ \hline
\multicolumn{1}{|c|}{\multirow{3}{*}{GP}}                                   & Accuracy  & .877$\pm$.00          & .840$\pm$.00          & .847$\pm$.00          & .844$\pm$.00          & .855$\pm$.00          & .797$\pm$.00          \\
\multicolumn{1}{|c|}{}                                                      & Precision & \textbf{.803$\pm$.00} & .809$\pm$.00          & \textbf{.697$\pm$.00} & \textbf{.854$\pm$.00} & \textbf{.883$\pm$.00} & \textbf{.934$\pm$.00} \\
\multicolumn{1}{|c|}{}                                                      & Recall    & \textbf{.958$\pm$.00} & .763$\pm$.00          & \textbf{.844$\pm$.00} & .860$\pm$.00          & .728$\pm$.00          & .718$\pm$.00          \\ \hline
\multicolumn{1}{|c|}{\multirow{3}{*}{GP-M (ours)}}                          & Accuracy  & \textbf{.884$\pm$.00} & \textbf{.961$\pm$.00} & \textbf{.932$\pm$.00} & \textbf{.947$\pm$.00} & \textbf{.965$\pm$.00} & \textbf{.953$\pm$.00} \\
\multicolumn{1}{|c|}{}                                                      & Precision & .768$\pm$.00          & .056$\pm$.00          & .341$\pm$.00          & .036$\pm$.00          & .299$\pm$.00          & .205$\pm$.00          \\
\multicolumn{1}{|c|}{}                                                      & Recall    & .677$\pm$.00          & \textbf{.999$\pm$.00} & .694$\pm$.00          & \textbf{.999$\pm$.00} & \textbf{.900$\pm$.00} & \textbf{.947$\pm$.00} \\ \hline
\end{tabular}
\end{center}
\label{tab: uncertainty estimation experiment result}
\end{table*}

\subsubsection{Uncertainty Estimation Accuracy}
In this experiment, we evaluated the uncertainty estimation accuracy of GP-M compared to other baseline methods. We integrated our fallback mechanism (described in Section \ref{Sec: fallback mechanism}) into all three models to measure their capability to accurately identify model errors exceeding $1^\circ C$. This threshold is stricter than the typical sensor fault tolerance of $2^\circ C$~\cite{kumar2021building}. We then assessed the accuracy, precision, and recall of each model over the following 30 days of the training dataset. To evaluate the stability of each method, we trained new models and repeated the experiment five times, calculating the mean and standard deviation of the results. These results are detailed in Table \ref{tab: uncertainty estimation experiment result}, with the best performances highlighted in bold for each metric across different environments.

We found that GP-M consistently outperformed both baselines in terms of overall accuracy. Notably, GP-M had a higher recall but lower precision compared to GP, indicating that while GP-M effectively identified the most significant model errors, it also mistakenly flagged some minor errors. This cautious approach ensures that GP-M minimizes the risk of overlooking critical errors larger than the threshold.

Regarding stability, both the standard GP and GP-M exhibited minimal instability, with variations less than $0.5\%$ across all metrics. In contrast, the DE model displayed considerable instability during the experiments, with a standard deviation reaching up to $12\%$ for recall. This instability is attributed to the random nature of parameter initialization and the inherent variability in the training process of neural network models. Conversely, GP-based methods showed a stronger resistance to such randomness, underscoring their reliability in uncertainty estimation.

\subsubsection{Model Convergence}
\begin{figure}
    \includegraphics[width=\columnwidth]{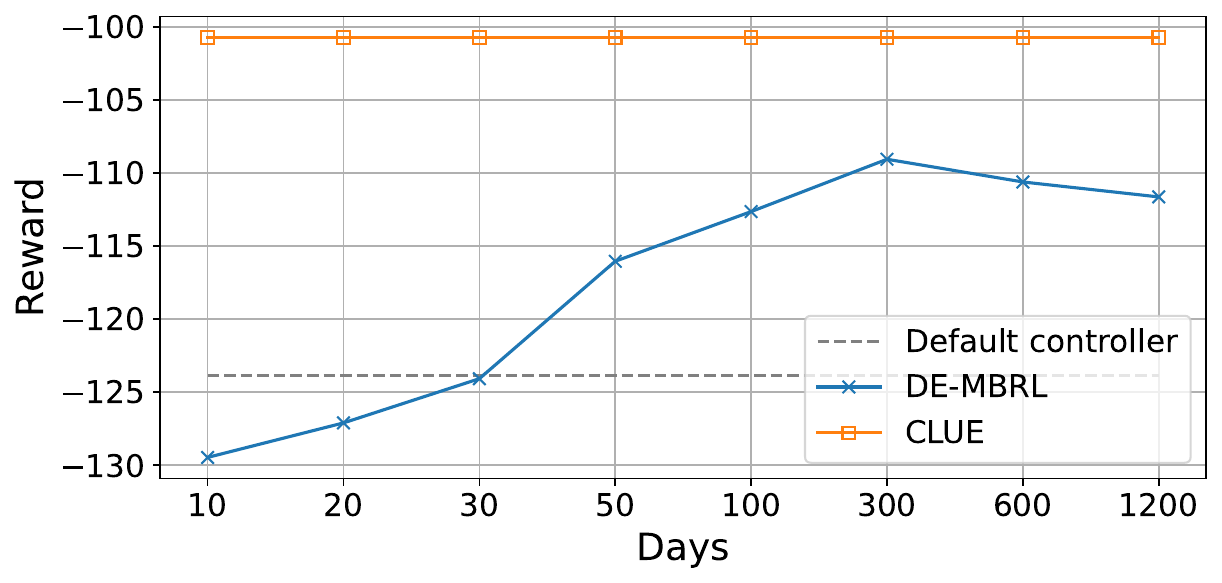}
    \caption{Data efficiency results: DE-MBRL vs. \textit{CLUE} convergence in cumulative reward. \textit{CLUE} stabilizes after 7 days, outperforming DE-MBRL.}
    \label{fig:reward_comparison}
\end{figure}

We conducted experiments to determine the convergence time step for \textit{CLUE} and its DE-MBRL counterpart in terms of cumulative reward, as depicted in Figure \ref{fig:reward_comparison}. In these tests, DE-MBRL was trained offline with varying amounts of data from the target building and subsequently used to control a 5-zone building in Pittsburgh during January. Conversely, \textit{CLUE} employed meta kernel learning followed by fine-tuning with different data volumes from the same building. The GP model was fitted to $700$ time steps of this building data (aligning with the data size used in \cite{goliatt2018modeling}) to balance effective modeling with computation efficiency. For setting the model error threshold $e^*$, we conducted an exhaustive search to determine the optimal threshold, ranging from $0.5^\circ C$ to $3^\circ C$.

The results showed that DE-MBRL required $50$ days of offline training data to exceed the performance of the default controller, and an additional $250$ days to achieve peak performance. In contrast, \textit{CLUE}'s performance stabilized after just $7$ days of data, with no further improvements in control performance with additional training. Based on this finding, we limited \textit{CLUE} to $7$ days of data for subsequent experiments.

For subsequent experiments, DE-MBRL was trained on $120,000$ time steps of data (approximately 3.42 years), collected using the default PID controller from the target building. \textit{CLUE} followed the same training regimen as initially outlined. The model error threshold was set at $0.5^\circ C$. Using these settings, we then evaluated \textit{CLUE} and baseline methods for controlling the building's HVAC system, focusing on thermal comfort and energy efficiency, which are discussed in the following subsections.

\subsection{Building Control}

\begin{figure}[t]
\includegraphics[width=\columnwidth]{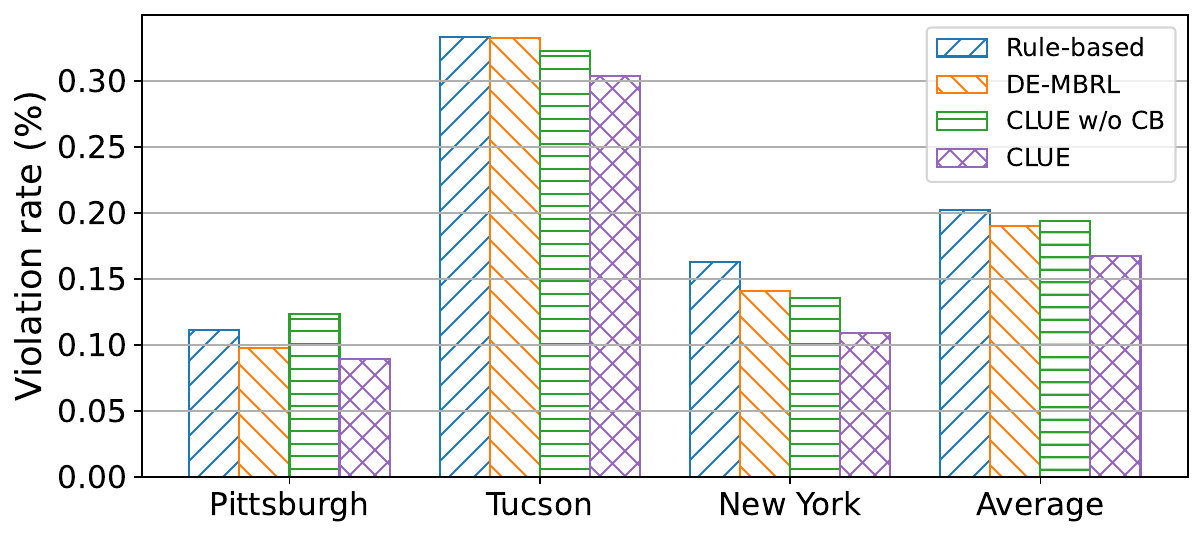}
\caption{Comfort violation rates: \textit{CLUE} vs. baselines across Pittsburgh, Tucson, and New York.}
\label{fig:violation_rate_comparison}
\end{figure}

\subsubsection{Thermal Comfort}
In our analysis, we quantitatively assessed the thermal comfort by examining the violation rates of various control methods, as depicted in Figure \ref{fig:violation_rate_comparison}. Our study reveals that \textit{CLUE} demonstrates superior performance over baseline methods across all tested locations—Pittsburgh, Tucson, and New York—as well as in the aggregated average. Notably, even when leveraging a GP-M dynamics model with a higher model error, as indicated by our model error comparison in Figure \ref{fig:compare_model_errors}, \textit{CLUE} secured lower violation rates than the DE-MBRL approach. This outcome underscores \textit{CLUE}'s robustness and its ability to generate effective control actions despite the presence of model inaccuracies. In contrast, the variant of \textit{CLUE} w/o CB underperformed relative to DE-MBRL. This disparity was anticipated and illustrates the significant role that CB plays in the efficacy of \textit{CLUE}. The inclusion of CB helps to mitigate the impact of model inaccuracies, highlighting the importance of uncertainty quantification in achieving high-quality control actions. Through this experiment in simulated environments that emulate diverse climatic conditions, \textit{CLUE} not only achieves a balance between energy efficiency and occupant comfort but also exhibits remarkable data efficiency. This is crucial for practical applications where extensive data collection is often challenging and costly.

\begin{figure}
\includegraphics[width=\columnwidth]{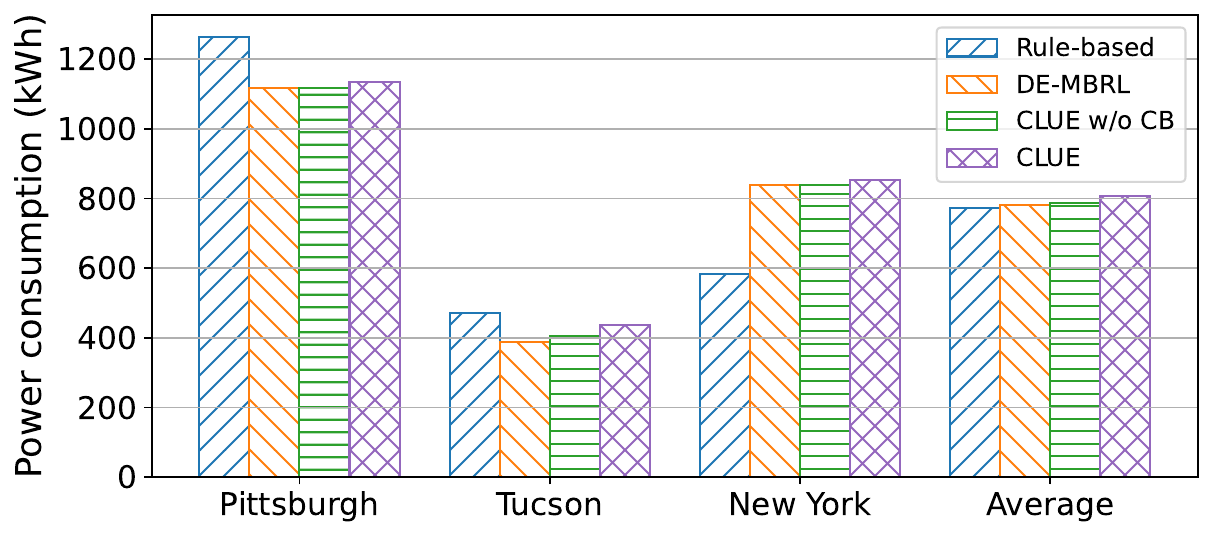}
\caption{Energy consumption: \textit{CLUE} vs. baseline methods}
\label{fig:power_consumption_comparison}
\end{figure}

\subsubsection{Energy Consumption}
Figure \ref{fig:power_consumption_comparison} compares the energy consumption of \textit{CLUE} with other established methods. Our analysis found that \textit{CLUE} tends to consume slightly more energy than methods without a confidence-based mechanism. This increase is linked to the fallback mechanism, which, when active, consumes energy comparable to a traditional rule-based system. Despite this, the energy use of \textit{CLUE} aligns well with other MBRL methods. Importantly, the energy \textit{CLUE} used during the fallback phase contributes to maintaining a high level of thermal comfort. Thus, \textit{CLUE} is more effective in providing comfortable conditions without significantly increasing energy use compared to the DE-MBRL approach. The advantage of \textit{CLUE} is its efficiency in offering more periods of comfort for every kilowatt-hour of energy used, making it a strong candidate for applications where occupant comfort is a critical factor. This efficient trade-off between energy consumption and comfort makes \textit{CLUE} particularly well-suited for use in settings that prioritize a comfortable environment. Its capability to sustain high comfort levels with a modest uptick in energy demand recommends it for use in locations like healthcare facilities, where comfort is essential, and residential areas that are looking to improve the well-being of residents while managing energy costs effectively.

\subsubsection{Effect of confidence-based control}
Our study showed \textit{CLUE}'s performance with its fallback mechanism against \textit{CLUE} without it (denoted as \textit{CLUE} w/o CB), as depicted in Figures \ref{fig:violation_rate_comparison} and \ref{fig:power_consumption_comparison}. \textit{CLUE} markedly reduced comfort violations compared to all baselines, demonstrating the fallback mechanism’s ability to handle model inaccuracies. Notably, \textit{CLUE} w/o CB outperformed the DE-MBRL model in two out of the three cities tested, suggesting that \textit{CLUE} can be effective even without the fallback under certain conditions. In terms of energy usage, \textit{CLUE} with the fallback mechanism showed a slight increase when compared to \textit{CLUE} w/o CB, attributable to the fallback's engagement during uncertainty, as shown in Figure \ref{fig:power_consumption_comparison}. Despite this, \textit{CLUE} remains efficient, balancing energy consumption with the assurance of maintained comfort levels.

The fallback mechanism's contribution to \textit{CLUE}’s robust performance across various environmental conditions is evident. It allows the system to manage uncertainty intelligently and maintain high control standards. The results support the incorporation of confidence measures into control strategies, significantly strengthening HVAC system performance by providing a calculated balance between energy efficiency and user comfort.

\begin{table*}
\caption{Translated uncertainty thresholds for various expected model error thresholds.}

\begin{center}
\begin{tabular}{|c|cc|cc|cc|}
\hline
\multirow{2}{*}{\begin{tabular}[c]{@{}c@{}}Expected model \\ error threshold ($^\circ C$)\end{tabular}} & \multicolumn{2}{c|}{Pittsburgh}                            & \multicolumn{2}{c|}{Tucson}                                & \multicolumn{2}{c|}{New York}                              \\ \cline{2-7} 
                                                                                                        & \multicolumn{1}{c|}{var threshold ($^\circ C$)} & accuracy & \multicolumn{1}{c|}{var threshold ($^\circ C$)} & accuracy & \multicolumn{1}{c|}{var threshold ($^\circ C$)} & accuracy \\ \hline
0.5                                                                                                     & \multicolumn{1}{c|}{0.3}                        & 0.869    & \multicolumn{1}{c|}{1.0}                        & 0.799    & \multicolumn{1}{c|}{0.5}                        & 0.889    \\
1.0                                                                                                     & \multicolumn{1}{c|}{0.8}                        & 0.885    & \multicolumn{1}{c|}{1.4}                        & 0.929    & \multicolumn{1}{c|}{0.9}                        & 0.966    \\
1.5                                                                                                     & \multicolumn{1}{c|}{1.2}                        & 0.885    & \multicolumn{1}{c|}{1.4}                        & 0.962    & \multicolumn{1}{c|}{1.7}                        & 0.981    \\
2.0                                                                                                     & \multicolumn{1}{c|}{2.4}                        & 0.902    & \multicolumn{1}{c|}{2.6}                        & 0.978    & \multicolumn{1}{c|}{1.8}                        & 0.992    \\
2.5                                                                                                     & \multicolumn{1}{c|}{3.0}                        & 0.911    & \multicolumn{1}{c|}{2.9}                        & 0.982    & \multicolumn{1}{c|}{2.8}                        & 0.995    \\
3.0                                                                                                     & \multicolumn{1}{c|}{3.5}                        & 0.918    & \multicolumn{1}{c|}{3.6}                        & 0.983    & \multicolumn{1}{c|}{2.8}                        & 0.996    \\ \hline
\end{tabular}
\end{center}
\label{tab: Translated uncertainty threshold}
\end{table*}

\begin{figure*}
\includegraphics[width=\textwidth]{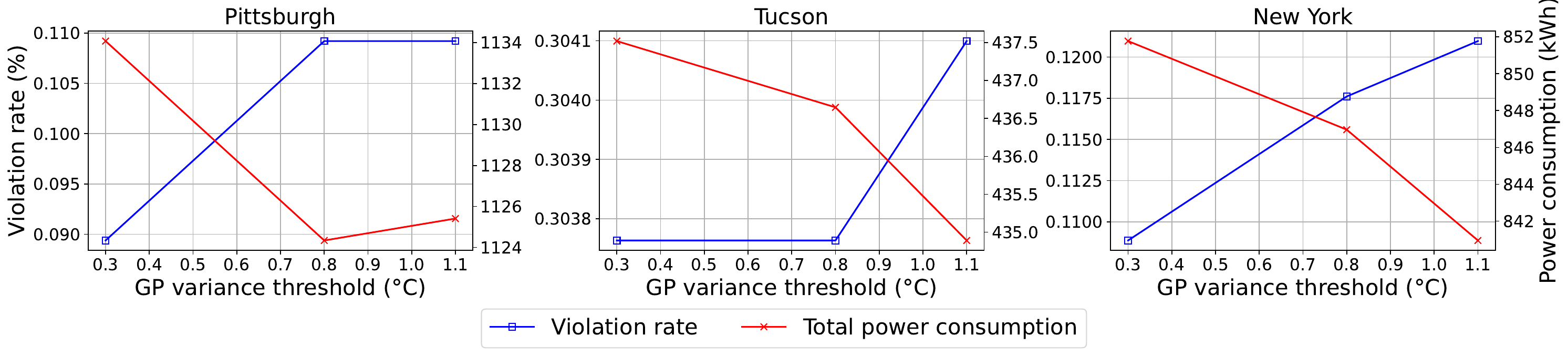}
\caption{Impact of GP variance thresholds on the control performance of \textit{CLUE} in buildings located in Pittsburgh, Tucson, and New York.}
\label{fig:violation_threshold}
\end{figure*}

\subsubsection{Effect of confidence threshold control knob}\label{Sec: Effect of confidence threshold control knob.}
This section examines how adjusting the confidence threshold—a key parameter in the confidence-based controller—affects the performance of \textit{CLUE}. We tested \textit{CLUE} with varying confidence thresholds in the same environments used in prior experiments to capture the threshold's impact. Our evaluation comprises two parts: first, assessing the empirical results from translating uncertainty thresholds (see Section \ref{Sec: Uncertainty Threshold Translation}), and second, observing how these translated thresholds influence control outcomes in simulated EnergyPlus environments.

\textbf{Empirical Results from Uncertainty Threshold Translation:}
Continuing from the previous experiments, after fine-tuning the GP kernel over $7$ days of data to obtain the thermal dynamics model, we utilized an uncertainty threshold translation module. This module interprets a set threshold for expected model error into corresponding uncertainty values produced by the GP model. To mimic real-world conditions where only a limited dataset is available, we employed the same $7$-day historical dataset, initially used for model fine-tuning, to facilitate the translation process. We processed a range of expected model error thresholds from $0.5^\circ C$ to $3^\circ C$, incrementing in steps of $0.5^\circ C$. For each increment, the translation module identified a GP variance threshold that maximized classification accuracy for identifying the predetermined model error. These translation accuracies are summarized in Table \ref{tab: Translated uncertainty threshold}.

A notable observation was the positive correlation between the GP variance thresholds and the expected model error thresholds. Additionally, higher thresholds yielded better classification accuracy. This increase in accuracy aligns with the skewed distribution of the $7$-day dataset, where sample counts dwindle at elevated error thresholds, leading to a smaller validation set and, consequently, an apparently higher accuracy.

\textbf{Impact of Varying Thresholds on Control Performance:}
Our investigation into the impact of different GP variance thresholds on the control performance of \textit{CLUE} utilized three specific values: 0.3, 0.8, and 1.1. These thresholds were applied in simulations of \textit{CLUE} managing buildings located in Pittsburgh, Tucson, and New York, as shown in Figure \ref{fig:violation_threshold}. A clear inverse relationship was observed: lower GP variance thresholds, indicating tighter control, corresponded with a decrease in comfort violations but an uptick in energy consumption. Conversely, higher thresholds, suggesting a more relaxed control approach, resulted in increased comfort violations but reduced energy usage.

This inverse trend was consistently observed across all three cities, indicating a robust pattern. In Pittsburgh, the violation rate saw a significant decrease from approximately 0.11\% to under 0.095\%, while energy consumption rose from about 1124 kWh to nearly 1134 kWh as the threshold tightened. Tucson displayed the most dramatic shift with a reduction in violation rates, accompanied by the sharpest increase in energy demand. In New York, the adjustment in thresholds also resulted in a marked improvement in comfort at the cost of higher energy requirements. The consistency of these trends across varied climates suggests that \textit{CLUE}'s fallback mechanism is responsive to the threshold settings, enabling a tunable balance between energy efficiency and thermal comfort.

\begin{figure}[t]
\includegraphics[width=\columnwidth]{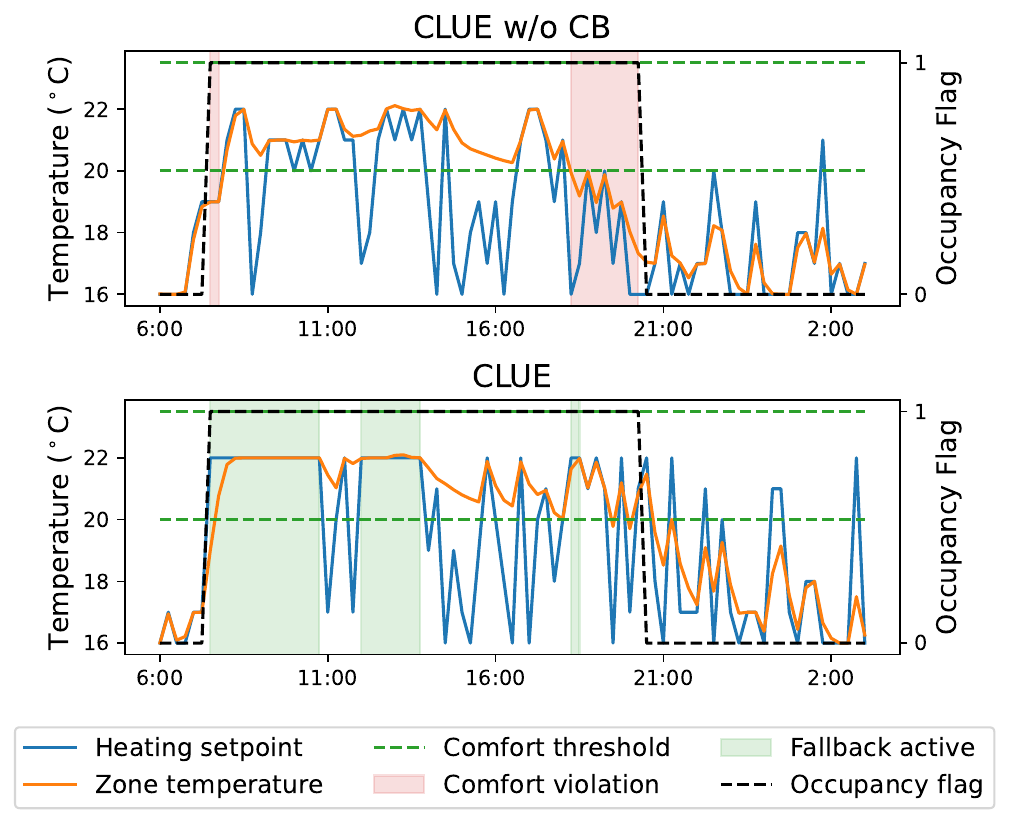}
\caption{Analysis of \textit{CLUE}'s control performance gain and fallback mechanism effectiveness.}
\label{fig:action_example}
\end{figure}

\subsubsection{Analysis of the control performance gain of \textit{CLUE}}
To investigate the reasons behind the building control performance gain of \textit{CLUE} compared with the baseline method, we plotted the data from one day of our simulation experiment for \textit{CLUE} and \textit{CLUE} w/o CB in Figure \ref{fig:action_example}. The simulation scenario was in January, so only the heating temperature setpoint was displayed. Out of the $85$ time steps shown in the figure, $26$ time steps had the fallback mechanism activated, i.e. the control action was overridden by the default controller about $30\%$ of the times. During a one-day period, an ideal controller is expected to keep the zone temperature within the comfort range during the occupied times and let the zone naturally cool down to the environment temperature during the unoccupied times to save energy. There are usually two origins of performance gain for MBRL approaches \cite{chen2019gnu}. First, the MBRL controllers learn to pre-heat the room to desired temperatures before the room is occupied to get a lower violation rate. Second, the MBRL approaches cool and re-heat the room repeatedly, keeping the temperature within the comfort range while saving energy during the time periods when the heating is turned off.

We found that the \textit{CLUE} w/o CB made a mistake mid-way between $16:00$ to $21:00$. It falsely believed that the zone temperature would take longer to cool down and would turn the heating off prematurely. This usually happens when the dynamics model overestimates the room's thermal capacity. The same mistake happened to the DE-MBRL controller. After observing that the temperature has plummeted, it underestimated the amount of heating to reheat the zone and eventually caused $135$ minutes of comfort violation. Our method, on the other hand, detected high model uncertainty and overrode $2$ time steps of control action with the default controller between $16:00$ to $21:00$. This successfully kept the zone temperature within the comfort range and also within the data distribution, allowing the controller to correctly predict the amount of heating needed for the rest of the occupied times. As a result, the fallback mechanism prevented $135$ minutes of comfort violation.

\begin{figure}
\includegraphics[width=\columnwidth]{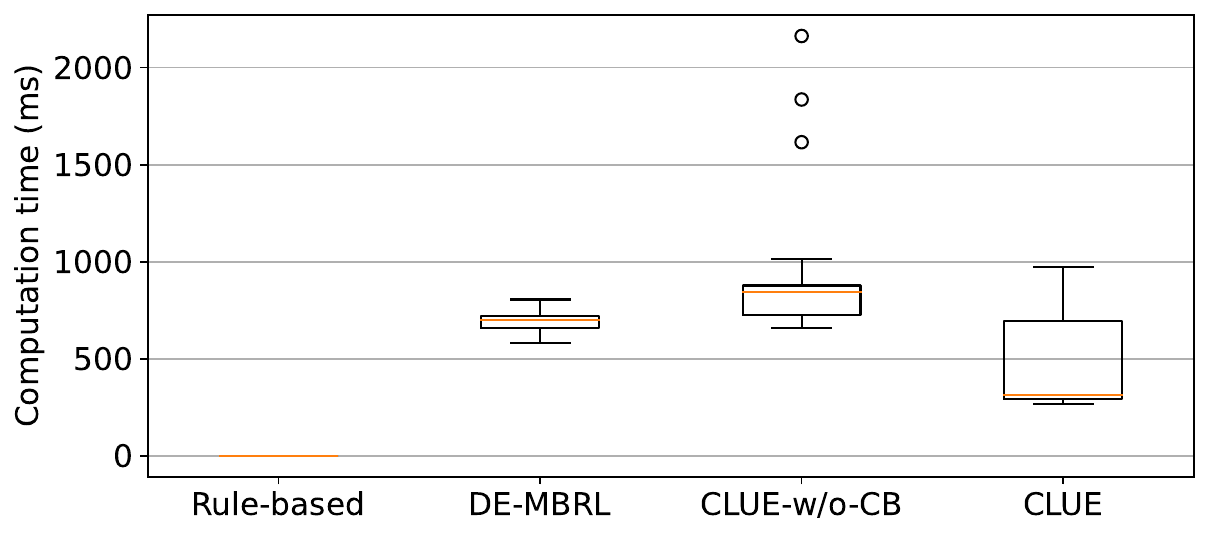}
\caption{Computation time analysis of \textit{CLUE} and baseline controllers.}
\label{fig:overhead}
\end{figure}

\subsubsection{Computation Overhead Analysis}
Figure \ref{fig:overhead} presents the computation time analysis for each controller utilized during our experimental evaluation. We observed that \textit{CLUE} w/o CB exhibited the highest computational overhead, with decision times generally spanning from 1 to 1.5 seconds. This extra time requirement is primarily due to the GP predictions, which tend to be more resource-intensive compared to the DE forward passes used in DE-MBRL.  In contrast, \textit{CLUE} demonstrates computational efficiency by occasionally bypassing complete trajectory rollouts. It utilizes the detection of high uncertainty in the immediate next time step to reduce the need for full prediction sequences, thus reducing processing times significantly. The rule-based controller, unsurprisingly, registered the lowest overhead, reflecting its less complex decision-making process. The comparable computational overheads of DE-MBRL and \textit{CLUE} w/o CB indicate that, although both methods utilize trajectory rollouts, the intensive computations required for the GP predictions in \textit{CLUE} w/o CB lead to longer processing times. The capability of \textit{CLUE} to streamline decision-making without sacrificing prediction accuracy or control efficacy illustrates an advanced optimization of computational resources, aligning well with the real-time demands of building control systems. This optimization highlights \textit{CLUE}'s practical application advantage where processing efficiency is as critical as control performance.

\section{Discussion}\label{sec: discussion}
Despite the promising results, there are several limitations and challenges to our approach that should be addressed in future research:

\begin{itemize}
    \item \textbf{Reliance on GP Model Accuracy:} The effectiveness of the GP model in predicting building dynamics heavily depends on the quality and quantity of the training data. In scenarios where the data is sparse or noisy, the model's accuracy can significantly degrade. While our meta-kernel learning approach enhances the initial parameter estimation, future work could explore advanced data augmentation techniques to synthetically expand the dataset or transfer learning methods to leverage pre-trained models from similar buildings. Furthermore, building models rarely remain static due to factors such as HVAC degradation and changes in building usage patterns. This necessitates the continuous updating of the model to maintain accuracy. Future research should focus on developing adaptive learning mechanisms that can dynamically update the GP model in response to changing building conditions, ensuring sustained performance over time.
    
    \item \textbf{Computational Overhead:} The GP predictions, especially during real-time operations, incur substantial computational costs. This overhead can hinder the scalability of the system for large-scale deployments involving multiple buildings. Future research could focus on developing more efficient GP approximations, such as sparse GP or leveraging variational inference techniques. Hybrid models that combine the strengths of GPs with faster machine learning techniques like deep learning could also be investigated to balance accuracy with computational efficiency. Additionally, incorporating the MPPI algorithm into the system, while effective, adds to the computational burden. Reporting the computational time of the proposed approach is crucial for understanding its practical applicability. Future work should include detailed computational time analyses and optimizations, such as parallel processing or leveraging specialized hardware (e.g., GPUs) to reduce the computational load and make the system more feasible for real-time applications.
    
    \item \textbf{Implementation Challenges:} Several challenges were encountered during the implementation of the CLUE system:
    \begin{itemize}
        \item \textbf{Initial Setup and Calibration:} Adapting CLUE to specific building characteristics requires extensive data collection and fine-tuning. This process can be resource-intensive and time-consuming. Automating parts of this process or developing more generalized models that require minimal fine-tuning could alleviate these challenges.
        \item \textbf{Simulation Limitations:} Our current focus was on simulation-based evaluations, and we did not integrate CLUE with various HVAC hardware and control protocols. Future research should explore real-world integrations and address compatibility issues with existing HVAC systems.
        \item \textbf{Ongoing Maintenance:} Regular maintenance and periodic retraining of the model are necessary to adapt to changing building dynamics and external conditions. Establishing automated monitoring and maintenance routines can ensure sustained performance over time.
    \end{itemize}
    
    \item \textbf{Practical Implications:} Deploying CLUE in real-world HVAC systems presents challenges:
    \begin{itemize}
        \item \textbf{Scalability:} Leveraging edge computing or cloud-based solutions to manage computational requirements can facilitate scalability across multiple buildings and systems. These approaches can distribute the computational load and provide real-time processing capabilities.
        \item \textbf{Integration:} Developing standardized protocols for integration with various HVAC hardware and control systems is essential. This ensures seamless operation and reduces the complexity of deploying CLUE in different buildings.
    \end{itemize}
\end{itemize}

Addressing these challenges through targeted research and development can pave the way for the successful deployment of CLUE in real-world settings, offering significant benefits in terms of energy efficiency and occupant comfort.

\section{Conclusion}\label{sec: conclusion}
This paper introduces \textit{CLUE}, a Model-Based Reinforcement Learning method specifically designed for HVAC control in buildings, focusing on enhancing energy efficiency while improving human comfort. By employing a GP to model building dynamics, \textit{CLUE} effectively integrates output uncertainty to increase the reliability of HVAC operations. The key innovation of our approach is a meta-kernel learning technique that utilizes domain knowledge from historical data of multiple buildings, significantly reducing the data required for GP hyperparameter tuning. This advanced technique is incorporated into an MPPI framework, enabling the system to compute optimal HVAC actions that prioritize energy efficiency and minimize risks associated with uncertain conditions.

Our empirical evaluation conducted in a simulated five-zone building demonstrates that \textit{CLUE} requires just seven days of training data to achieve performance levels that meet or exceed those of existing state-of-the-art MBRL methods. Notably, it accomplishes this with a 12.07\% reduction in comfort violations. These results underscore \textit{CLUE}’s capability as a scalable and efficient solution for HVAC control, offering significant advancements in both energy conservation and occupant comfort. This positions \textit{CLUE} as a potent tool for sustainable building management, potentially setting new standards for the integration of MBRL in real-world applications.

\bibliographystyle{IEEEtran}
\bibliography{cas-refs}

\begin{thebibliography}{10}
\providecommand{\url}[1]{#1}
\csname url@samestyle\endcsname
\providecommand{\newblock}{\relax}
\providecommand{\bibinfo}[2]{#2}
\providecommand{\BIBentrySTDinterwordspacing}{\spaceskip=0pt\relax}
\providecommand{\BIBentryALTinterwordstretchfactor}{4}
\providecommand{\BIBentryALTinterwordspacing}{\spaceskip=\fontdimen2\font plus
\BIBentryALTinterwordstretchfactor\fontdimen3\font minus \fontdimen4\font\relax}
\providecommand{\BIBforeignlanguage}[2]{{%
\expandafter\ifx\csname l@#1\endcsname\relax
\typeout{** WARNING: IEEEtran.bst: No hyphenation pattern has been}%
\typeout{** loaded for the language `#1'. Using the pattern for}%
\typeout{** the default language instead.}%
\else
\language=\csname l@#1\endcsname
\fi
#2}}
\providecommand{\BIBdecl}{\relax}
\BIBdecl

\bibitem{an2023clue}
Z.~An, X.~Ding, A.~Rathee, and W.~Du, ``Clue: Safe model-based rl hvac control using epistemic uncertainty estimation,'' in \emph{Proceedings of the 10th ACM International Conference on Systems for Energy-Efficient Buildings, Cities, and Transportation}, 2023, pp. 149--158.

\bibitem{ding2019octopus}
X.~Ding, W.~Du, and A.~Cerpa, ``Octopus: Deep reinforcement learning for holistic smart building control,'' in \emph{Proceedings of the 6th ACM international conference on systems for energy-efficient buildings, cities, and transportation}, 2019, pp. 326--335.

\bibitem{zhang2018practical}
Z.~Zhang and K.~P. Lam, ``Practical implementation and evaluation of deep reinforcement learning control for a radiant heating system,'' in \emph{Proceedings of the 5th Conference on Systems for Built Environments}, 2018, pp. 148--157.

\bibitem{park2020hvaclearn}
J.~Y. Park and Z.~Nagy, ``Hvaclearn: A reinforcement learning based occupant-centric control for thermostat set-points,'' in \emph{Proceedings of the Eleventh ACM International Conference on Future Energy Systems}, 2020, pp. 434--437.

\bibitem{moerland2023model}
T.~M. Moerland \emph{et~al.}, ``Model-based reinforcement learning: A survey,'' \emph{Foundations and Trends{\textregistered} in Machine Learning}, vol.~16, no.~1, pp. 1--118, 2023.

\bibitem{agbi2012parameter}
C.~Agbi, Z.~Song, and B.~Krogh, ``Parameter identifiability for multi-zone building models,'' in \emph{2012 IEEE 51st IEEE conference on decision and control (CDC)}.\hskip 1em plus 0.5em minus 0.4em\relax IEEE, 2012, pp. 6951--6956.

\bibitem{ding2020mb2c}
X.~Ding, W.~Du, and A.~E. Cerpa, ``Mb2c: Model-based deep reinforcement learning for multi-zone building control,'' in \emph{Proceedings of the 7th ACM international conference on systems for energy-efficient buildings, cities, and transportation}, 2020, pp. 50--59.

\bibitem{chen2019gnu}
B.~Chen, Z.~Cai, and M.~Berg{\'e}s, ``Gnu-rl: A precocial reinforcement learning solution for building hvac control using a differentiable mpc policy,'' in \emph{Proceedings of the 6th ACM international conference on systems for energy-efficient buildings, cities, and transportation}, 2019, pp. 316--325.

\bibitem{ding2024multi}
X.~Ding, A.~Cerpa, and W.~Du, ``Multi-zone hvac control with model-based deep reinforcement learning,'' \emph{IEEE Transactions on Automation Science and Engineering}, 2024.

\bibitem{baek2023uncertainty}
W.-J. Baek \emph{et~al.}, ``Uncertainty estimation for safe human-robot collaboration using conservation measures,'' in \emph{Proceedings of IAS-17}, 2023.

\bibitem{chua2018deep}
K.~Chua, R.~Calandra, R.~McAllister, and S.~Levine, ``Deep reinforcement learning in a handful of trials using probabilistic dynamics models,'' \emph{Advances in neural information processing systems}, vol.~31, 2018.

\bibitem{vermorel2005multi}
J.~Vermorel and M.~Mohri, ``Multi-armed bandit algorithms and empirical evaluation,'' in \emph{European conference on machine learning}.\hskip 1em plus 0.5em minus 0.4em\relax Springer, 2005.

\bibitem{tang2017exploration}
H.~Tang, R.~Houthooft, D.~Foote, A.~Stooke, O.~Xi~Chen, Y.~Duan, J.~Schulman, F.~DeTurck, and P.~Abbeel, ``\# exploration: A study of count-based exploration for deep reinforcement learning,'' \emph{Advances in neural information processing systems}, vol.~30, 2017.

\bibitem{lakshminarayanan2017simple}
B.~Lakshminarayanan, A.~Pritzel, and C.~Blundell, ``Simple and scalable predictive uncertainty estimation using deep ensembles,'' \emph{Advances in neural information processing systems}, vol.~30, 2017.

\bibitem{MASSAGRAY2016119}
F.~{Massa Gray} and M.~Schmidt, ``Thermal building modelling using gaussian processes,'' \emph{Energy and Buildings}, vol. 119, pp. 119--128, 2016.

\bibitem{goliatt2018modeling}
L.~Goliatt, P.~Capriles, and G.~R. Duarte, ``Modeling heating and cooling loads in buildings using gaussian processes,'' in \emph{2018 IEEE Congress on Evolutionary Computation (CEC)}.\hskip 1em plus 0.5em minus 0.4em\relax IEEE, 2018, pp. 1--6.

\bibitem{nghiem2017data}
T.~X. Nghiem and C.~N. Jones, ``Data-driven demand response modeling and control of buildings with gaussian processes,'' in \emph{2017 American Control Conference (ACC)}.\hskip 1em plus 0.5em minus 0.4em\relax IEEE, 2017, pp. 2919--2924.

\bibitem{duvenaud2014automatic}
D.~Duvenaud, ``Automatic model construction with gaussian processes,'' Ph.D. dissertation, University of Cambridge, 2014.

\bibitem{doe2010energyplus}
DoE, ``Energyplus input output reference,'' \emph{US Department of Energy}, 2010.

\bibitem{beltran2014optimal}
A.~Beltran and A.~E. Cerpa, ``Optimal hvac building control with occupancy prediction,'' in \emph{Proceedings of the 1st ACM conference on embedded systems for energy-efficient buildings}, 2014, pp. 168--171.

\bibitem{li2023economic}
H.~Li, J.~Xu, Q.~Zhao, and S.~Wang, ``Economic model predictive control in buildings based on piecewise linear approximation of predicted mean vote index,'' \emph{IEEE Transactions on Automation Science and Engineering}, 2023.

\bibitem{winkler2020office}
D.~A. Winkler, A.~Yadav, C.~Chitu, and A.~E. Cerpa, ``Office: Optimization framework for improved comfort \& efficiency,'' in \emph{2020 19th ACM/IEEE International Conference on Information Processing in Sensor Networks (IPSN)}.\hskip 1em plus 0.5em minus 0.4em\relax IEEE, 2020, pp. 265--276.

\bibitem{vazquez2020marlisa}
J.~R. Vazquez-Canteli, G.~Henze, and Z.~Nagy, ``Marlisa: Multi-agent reinforcement learning with iterative sequential action selection for load shaping of grid-interactive connected buildings,'' in \emph{Proceedings of the 7th ACM International Conference on Systems for Energy-Efficient Buildings, Cities, and Transportation}, 2020.

\bibitem{gao2020deepcomfort}
G.~Gao, J.~Li, and Y.~Wen, ``Deepcomfort: Energy-efficient thermal comfort control in buildings via reinforcement learning,'' \emph{IEEE Internet of Things Journal}, 2020.

\bibitem{lei2022practical}
Y.~Lei, S.~Zhan, E.~Ono, Y.~Peng, Z.~Zhang, T.~Hasama, and A.~Chong, ``A practical deep reinforcement learning framework for multivariate occupant-centric control in buildings,'' \emph{Applied Energy}, vol. 324, p. 119742, 2022.

\bibitem{chen2020gnu}
B.~Chen, Z.~Cai, and M.~Berg{\'e}s, ``Gnu-rl: A practical and scalable reinforcement learning solution for building hvac control using a differentiable mpc policy,'' \emph{Frontiers in Built Environment}, vol.~6, p. 562239, 2020.

\bibitem{zhang2019building}
C.~Zhang, S.~R. Kuppannagari, R.~Kannan, and V.~K. Prasanna, ``Building hvac scheduling using reinforcement learning via neural network based model approximation,'' in \emph{Proceedings of the 6th ACM international conference on systems for energy-efficient buildings, cities, and transportation}, 2019, pp. 287--296.

\bibitem{chen2022mbrl}
L.~Chen, F.~Meng, and Y.~Zhang, ``Mbrl-mc: An hvac control approach via combining model-based deep reinforcement learning and model predictive control,'' \emph{IEEE Internet of Things Journal}, vol.~9, no.~19, pp. 19\,160--19\,173, 2022.

\bibitem{baldi2018automating}
S.~Baldi, C.~D. Korkas, M.~Lv, and E.~B. Kosmatopoulos, ``Automating occupant-building interaction via smart zoning of thermostatic loads: A switched self-tuning approach,'' \emph{Applied energy}, vol. 231, pp. 1246--1258, 2018.

\bibitem{korkas2018grid}
C.~D. Korkas, S.~Baldi, and E.~B. Kosmatopoulos, ``Grid-connected microgrids: Demand management via distributed control and human-in-the-loop optimization,'' in \emph{Advances in renewable energies and power technologies}.\hskip 1em plus 0.5em minus 0.4em\relax Elsevier, 2018, pp. 315--344.

\bibitem{michailidis2015proactive}
I.~T. Michailidis, S.~Baldi, M.~F. Pichler, E.~B. Kosmatopoulos, and J.~R. Santiago, ``Proactive control for solar energy exploitation: A german high-inertia building case study,'' \emph{Applied Energy}, vol. 155, pp. 409--420, 2015.

\bibitem{baldi2015model}
S.~Baldi, I.~Michailidis, C.~Ravanis, and E.~B. Kosmatopoulos, ``Model-based and model-free “plug-and-play” building energy efficient control,'' \emph{Applied Energy}, vol. 154, pp. 829--841, 2015.

\bibitem{an2024go}
Z.~An, X.~Ding, and W.~Du, ``Go beyond black-box policies: Rethinking the design of learning agent for interpretable and verifiable hvac control,'' \emph{arXiv preprint arXiv:2403.00172}, 2024.

\bibitem{jayant2022model}
A.~K. Jayant and S.~Bhatnagar, ``Model-based safe deep reinforcement learning via a constrained proximal policy optimization algorithm,'' \emph{Advances in Neural Information Processing Systems}, vol.~35, pp. 24\,432--24\,445, 2022.

\bibitem{liu2022safe}
H.-Y. Liu, B.~Balaji, S.~Gao, R.~Gupta, and D.~Hong, ``Safe hvac control via batch reinforcement learning,'' in \emph{2022 ACM/IEEE 13th International Conference on Cyber-Physical Systems (ICCPS)}.\hskip 1em plus 0.5em minus 0.4em\relax IEEE, 2022, pp. 181--192.

\bibitem{zhang2022safe}
C.~Zhang, S.~R. Kuppannagari, and V.~K. Prasanna, ``Safe building hvac control via batch reinforcement learning,'' \emph{IEEE Transactions on Sustainable Computing}, 2022.

\bibitem{buonomano2015adaptive}
A.~Buonomano, U.~Montanaro, A.~Palombo, and S.~Santini, ``Adaptive control for building thermo-hygrometric analysis: a novel dynamic simulation code for indoor spaces with multi-enclosed thermal zones,'' \emph{Energy Procedia}, vol.~78, pp. 2190--2195, 2015.

\bibitem{short2012real}
M.~Short, ``Real-time infinite horizon adaptive/predictive control for smart home hvac applications,'' in \emph{Proceedings of 2012 IEEE 17th international conference on emerging technologies \& factory automation (ETFA 2012)}.\hskip 1em plus 0.5em minus 0.4em\relax IEEE, 2012, pp. 1--8.

\bibitem{yang2019adaptive}
S.~Yang, M.~P. Wan, W.~Chen, B.~F. Ng, and D.~Zhai, ``An adaptive robust model predictive control for indoor climate optimization and uncertainties handling in buildings,'' \emph{Building and Environment}, vol. 163, p. 106326, 2019.

\bibitem{schmelas2015adaptive}
M.~Schmelas, T.~Feldmann, and E.~Bollin, ``Adaptive predictive control of thermo-active building systems (tabs) based on a multiple regression algorithm,'' \emph{Energy and Buildings}, vol. 103, pp. 14--28, 2015.

\bibitem{tesfay2018adaptive}
M.~Tesfay, F.~Alsaleem, P.~Arunasalam, and A.~Rao, ``Adaptive-model predictive control of electronic expansion valves with adjustable setpoint for evaporator superheat minimization,'' \emph{Building and Environment}, vol. 133, pp. 151--160, 2018.

\bibitem{tanaskovic2017robust}
M.~Tanaskovic, D.~Sturzenegger, R.~Smith, and M.~Morari, ``Robust adaptive model predictive building climate control,'' \emph{Ifac-Papersonline}, vol.~50, no.~1, pp. 1871--1876, 2017.

\bibitem{jia2018event}
Q.-S. Jia, J.~Wu, Z.~Wu, and X.~Guan, ``Event-based hvac control—a complexity-based approach,'' \emph{IEEE Transactions on Automation Science and Engineering}, vol.~15, no.~4, pp. 1909--1919, 2018.

\bibitem{patti2014event}
E.~Patti, A.~Acquaviva, M.~Jahn, F.~Pramudianto, R.~Tomasi, D.~Rabourdin, J.~Virgone, and E.~Macii, ``Event-driven user-centric middleware for energy-efficient buildings and public spaces,'' \emph{IEEE Systems Journal}, vol.~10, no.~3, pp. 1137--1146, 2014.

\bibitem{wu2015optimal}
Z.~Wu, Q.-S. Jia, and X.~Guan, ``Optimal control of multiroom hvac system: An event-based approach,'' \emph{IEEE Transactions on Control Systems Technology}, vol.~24, no.~2, pp. 662--669, 2015.

\bibitem{dhar2017adaptive}
N.~K. Dhar, N.~K. Verma, and L.~Behera, ``Adaptive critic-based event-triggered control for hvac system,'' \emph{IEEE Transactions on Industrial Informatics}, vol.~14, no.~1, pp. 178--188, 2017.

\bibitem{sun2015event}
B.~Sun, P.~B. Luh, Q.-S. Jia, and B.~Yan, ``Event-based optimization within the lagrangian relaxation framework for energy savings in hvac systems,'' \emph{IEEE Transactions on Automation Science and Engineering}, vol.~12, no.~4, pp. 1396--1406, 2015.

\bibitem{wang2016event}
J.~Wang, G.~Huang, Y.~Sun, and X.~Liu, ``Event-driven optimization of complex hvac systems,'' \emph{Energy and Buildings}, vol. 133, pp. 79--87, 2016.

\bibitem{wang2018event}
J.~Wang, Q.-S. Jia, G.~Huang, and Y.~Sun, ``Event-driven optimal control of central air-conditioning systems: Event-space establishment,'' \emph{Science and Technology for the Built Environment}, vol.~24, no.~8, pp. 839--849, 2018.

\bibitem{rockett2017model}
P.~Rockett and E.~A. Hathway, ``Model-predictive control for non-domestic buildings: a critical review and prospects,'' \emph{Building Research \& Information}, vol.~45, no.~5, pp. 556--571, 2017.

\bibitem{killian2016ten}
M.~Killian and M.~Kozek, ``Ten questions concerning model predictive control for energy efficient buildings,'' \emph{Building and Environment}, vol. 105, pp. 403--412, 2016.

\bibitem{privara2013building}
S.~Privara, J.~Cigler, Z.~V{\'a}{\v{n}}a, F.~Oldewurtel, C.~Sagerschnig, and E.~{\v{Z}}{\'a}{\v{c}}ekov{\'a}, ``Building modeling as a crucial part for building predictive control,'' \emph{Energy and Buildings}, vol.~56, pp. 8--22, 2013.

\bibitem{lu2015modeling}
X.~L{\"u}, T.~Lu, C.~J. Kibert, and M.~Viljanen, ``Modeling and forecasting energy consumption for heterogeneous buildings using a physical--statistical approach,'' \emph{Applied Energy}, vol. 144, pp. 261--275, 2015.

\bibitem{le2021deep}
D.~V. Le, R.~Wang, Y.~Liu, R.~Tan, Y.-W. Wong, and Y.~Wen, ``Deep reinforcement learning for tropical air free-cooled data center control,'' \emph{ACM Transactions on Sensor Networks (TOSN)}, 2021.

\bibitem{ding2024exploring}
X.~Ding, A.~Cerpa, and W.~Du, ``Exploring deep reinforcement learning for holistic smart building control,'' \emph{ACM Transactions on Sensor Networks}, vol.~20, no.~3, pp. 1--28, 2024.

\bibitem{ding2022drlic}
X.~Ding and W.~Du, ``Drlic: Deep reinforcement learning for irrigation control,'' in \emph{2022 21st ACM/IEEE International Conference on Information Processing in Sensor Networks (IPSN)}.\hskip 1em plus 0.5em minus 0.4em\relax IEEE, 2022, pp. 41--53.

\bibitem{an2024reward}
Z.~An, X.~Ding, and W.~Du, ``Reward bound for behavioral guarantee of model-based planning agents,'' \emph{arXiv preprint arXiv:2402.13419}, 2024.

\bibitem{achiam2017constrained}
J.~Achiam, D.~Held, A.~Tamar, and P.~Abbeel, ``Constrained policy optimization,'' in \emph{International conference on machine learning}.\hskip 1em plus 0.5em minus 0.4em\relax PMLR, 2017, pp. 22--31.

\bibitem{zhang2018deep}
Z.~Zhang \emph{et~al.}, ``A deep reinforcement learning approach to using whole building energy model for hvac optimal control,'' in \emph{2018 Building Performance Analysis Conference and SimBuild}, vol.~3, 2018, pp. 22--23.

\bibitem{du2023optimizing}
X.~Ding and W.~Du, ``Optimizing irrigation efficiency using deep reinforcement learning in the field,'' \emph{ACM Transactions on Sensor Networks}, 2024.

\bibitem{jimenez2021sinergym}
J.~Jim{\'e}nez-Raboso \emph{et~al.}, ``Sinergym: a building simulation and control framework for training reinforcement learning agents,'' in \emph{ACM BuildSys}, 2021, pp. 319--323.

\bibitem{modes}
H.~Rajabi, Z.~Hu, X.~Ding, S.~Pan, W.~Du, and A.~Cerpa, ``Modes: M ulti-sensor o ccupancy d ata-driven e stimation s ystem for smart buildings,'' in \emph{Proceedings of the Thirteenth ACM International Conference on Future Energy Systems}, 2022, pp. 228--239.

\bibitem{TODOS}
H.~Rajabi, X.~Ding, W.~Du, and A.~Cerpa, ``Todos: Thermal sensor data-driven occupancy estimation system for smart buildings,'' in \emph{Proceedings of the 10th ACM international conference on systems for energy-efficient buildings, cities, and transportation}, 2023, pp. 198--207.

\bibitem{kumar2021building}
D.~Kumar, X.~Ding, W.~Du, and A.~Cerpa, ``Building sensor fault detection and diagnostic system,'' in \emph{Proceedings of the 8th ACM International Conference on Systems for Energy-Efficient Buildings, Cities, and Transportation}, 2021, pp. 357--360.

\bibitem{finn2017model}
C.~Finn, P.~Abbeel, and S.~Levine, ``Model-agnostic meta-learning for fast adaptation of deep networks,'' in \emph{International conference on machine learning}.\hskip 1em plus 0.5em minus 0.4em\relax PMLR, 2017, pp. 1126--1135.

\bibitem{williams2016aggressive}
G.~Williams, P.~Drews, B.~Goldfain, J.~M. Rehg, and E.~A. Theodorou, ``Aggressive driving with model predictive path integral control,'' in \emph{2016 IEEE International Conference on Robotics and Automation (ICRA)}.\hskip 1em plus 0.5em minus 0.4em\relax IEEE, 2016, pp. 1433--1440.

\bibitem{paszke2019pytorch}
A.~Paszke, S.~Gross, F.~Massa, A.~Lerer, J.~Bradbury, G.~Chanan, T.~Killeen, Z.~Lin, N.~Gimelshein, L.~Antiga \emph{et~al.}, ``Pytorch: An imperative style, high-performance deep learning library,'' \emph{Advances in neural information processing systems}, vol.~32, 2019.

\bibitem{gardner2018gpytorch}
J.~Gardner, G.~Pleiss, K.~Q. Weinberger, D.~Bindel, and A.~G. Wilson, ``Gpytorch: Blackbox matrix-matrix gaussian process inference with gpu acceleration,'' \emph{Advances in neural information processing systems}, vol.~31, 2018.

\bibitem{standard2020ansi}
A.~STANDARD, ``Ansi/ashrae addendum a to ansi/ashrae standard 169-2020,'' \emph{ASHRAE Standing Standard Project Committee}, 2020.

\bibitem{hand2018note}
D.~Hand and P.~Christen, ``A note on using the f-measure for evaluating record linkage algorithms,'' \emph{Statistics and Computing}, vol.~28, pp. 539--547, 2018.

\end{thebibliography}

\end{document}